\documentclass[acmsmall]{acmart}

\usepackage{amsmath,amsfonts}

\usepackage{amssymb}
\usepackage{multirow}
\usepackage{algorithmic}
\newcommand{\rev}[1]{#1}
\usepackage{graphicx}
\usepackage{textcomp}
\usepackage{xcolor}
\usepackage{array}
\usepackage{siunitx}
\usepackage{url}
\usepackage{pifont}
\usepackage{subcaption}

\begin{document}

\setcopyright{cc}
\setcctype{by}
\acmJournal{POMACS}
\acmYear{2026} \acmVolume{10} \acmNumber{2} \acmArticle{28}
\acmMonth{6} \acmDOI{10.1145/3805626}

\title{ACALSim: A Scalable Parallel Simulation Framework for High-Performance System Design Space Exploration}

\author{Wei-Fen Lin}
\email{weifen@twhpcedu.org}
\affiliation{%
  \institution{Mijotech Inc.}
  \country{Taiwan}
}
\affiliation{%
  \institution{Taiwan High-Performance Computing Education Association}
  \country{Taiwan}
}

\author{Jen-Chien Chang}
\email{jenchien@twhpcedu.org}
\orcid{0000-0002-1814-8797}
\affiliation{%
  \institution{National Cheng Kung University}
  \city{Tainan}
  \country{Taiwan}
}
\affiliation{%
  \institution{Taiwan High-Performance Computing Education Association}
  \country{Taiwan}
}

\author{Yen-Po Chen}
\email{yenpo@twhpcedu.org}
\orcid{0000-0002-5932-454X}
\affiliation{%
  \institution{National Cheng Kung University}
  \city{Tainan}
  \country{Taiwan}
}
\affiliation{%
  \institution{Taiwan High-Performance Computing Education Association}
  \country{Taiwan}
}

\author{Zi-Yi Tai}
\email{elisatai@twhpcedu.org}
\affiliation{%
  \institution{Taiwan High-Performance Computing Education Association}
  \country{Taiwan}
}

\author{Yu-Cheng Chang}
\email{yuchang@twhpcedu.org}
\affiliation{%
  \institution{National Yang Ming Chiao Tung University}
  \country{Taiwan}
}
\affiliation{%
  \institution{Taiwan High-Performance Computing Education Association}
  \country{Taiwan}
}

\author{Chia-Pao Chiang}
\email{daniel100373@gmail.com}
\orcid{0009-0007-0295-4878}
\affiliation{%
  \institution{National Cheng Kung University}
  \city{Tainan}
  \country{Taiwan}
}

\author{Yu-Yang Lee}
\email{ppalee1217@gmail.com}
\orcid{0009-0000-1071-8524}
\affiliation{%
  \institution{National Cheng Kung University}
  \city{Tainan}
  \country{Taiwan}
}

\author{Yu-Jie Wang}
\email{pine0113@gmail.com}
\orcid{0009-0005-2649-6408}
\affiliation{%
  \institution{National Cheng Kung University}
  \city{Tainan}
  \country{Taiwan}
}

\renewcommand{\shortauthors}{Wei-Fen Lin et al.}

\begin{abstract}
Architectural simulation has become the critical bottleneck limiting design space exploration for high-performance computing systems. The growing complexity of modern GPUs and AI accelerators---with hundreds to thousands of tightly-coupled components---demands simulation frameworks that deliver efficient parallelism and scalable single-node execution. Existing frameworks such as SST, gem5, and GPGPU-Sim fail to meet these requirements: SST focuses on multi-node MPI scalability but struggles with intra-node scaling; GPGPU-Sim remains largely single-threaded. Critically, these frameworks provide fixed threading models with no mechanism for users to optimize simulation performance for their specific workloads.

We introduce ACALSim, a scalable parallel simulation framework designed to accelerate design space exploration for high-performance systems. As a \textit{framework} contribution, ACALSim provides infrastructure and APIs for building high-performance simulators---timing model accuracy is the responsibility of simulator developers, not the framework. ACALSim's key innovation is a \textbf{pluggable thread management architecture} that enables developers to implement custom scheduling strategies optimized for specific simulation patterns---a capability absent in existing frameworks. This is complemented by (1) event-driven execution with fast-forward to eliminate idle cycle overhead, (2) a shared-memory data model enabling zero-copy communication, and (3) a two-phase parallel execution model for deterministic thread scaling. We demonstrate ACALSim's effectiveness through HPCSim, a GPU simulator targeting A100-class architectures. Direct comparison with an SST implementation---using identical shared timing cores to isolate framework overhead---shows ACALSim achieves \textbf{over 14$\times$ speedup} with 41\% lower memory footprint, while hardware validation confirms 0.72--1.22$\times$ cycle count correlation with A100 measurements. While SST fails to complete 256+ thread block workloads within practical time limits, ACALSim simulates full LLaMA transformer layers (single block) in 17.7 minutes for LLaMA-7B and 30.4 minutes for LLaMA-13B---enabling practical design space exploration that SST cannot achieve.
\end{abstract}



\received{January 2026}
\received[revised]{March 2026}
\received[accepted]{April 2026}

\maketitle

\section{Introduction}

The rapid proliferation of large-scale AI models has fundamentally transformed the landscape of high-performance computing (HPC) system design. Modern AI accelerators---from GPUs with hundreds of streaming multiprocessors to custom AI chips with specialized tensor units---must be designed and validated within increasingly compressed product cycles. This paradigm shift necessitates early-stage architectural simulation to evaluate design trade-offs through comprehensive hardware-software co-simulation before committing to silicon. The development of effective solutions in this domain requires unprecedented collaboration among multiple research teams, each contributing specialized expertise to address both timing constraints and design complexity. However, the collaborative nature of modern AI computing system development, coupled with their inherent complexity, presents several significant challenges:

\begin{table}[t]
\centering
\caption{Comparison of Contemporary Computer Architecture Simulators}
\footnotesize
\begin{tabular}{|l|c|c|c|p{3.8cm}|}
\hline
\textbf{Simulator} & \textbf{Perf} & \textbf{PSS} & \textbf{OSS} & \textbf{Characteristics} \\
\hline
\multicolumn{5}{|c|}{\textit{Performance-Oriented Simulators}} \\
\hline
FireSim~\cite{firesim} & VH & Full & \ding{51} & +FPGA-accelerated; -HW dependent \\
\hline
GPGPU-Sim~\cite{gpgpusim} & H & Full & \ding{51} & +GPU modeling; -GPU specific \\
\hline
SST~\cite{sst} & M & Full & \ding{51} & +Component-based; -MPI overhead \\
\hline
ZSim~\cite{zsim} & H & Full & \ding{51} & +Many-core sim; -x86 specific \\
\hline
\multicolumn{5}{|c|}{\textit{System-Level Simulators}} \\
\hline
SystemC-TLM~\cite{systemc} & L & Ltd & \ding{51} & +TLM; -Limited parallelism \\
\hline
Gem5-SystemC~\cite{gem5} & L & Ltd & \ding{51} & +Cycle-accurate; -Overhead \\
\hline
QEMU-SystemC~\cite{qemu} & L & Ltd & \ding{51} & +Fast functional; -Low accuracy \\
\hline
Simics~\cite{simics} & H & Full & $\times$ & +Full-system; -Commercial \\
\hline
\multicolumn{5}{|p{7.5cm}|}{\scriptsize \textbf{Note:} Perf=Performance (VH/H/M/L), PSS=Parallel Support, OSS=Open-Source, Ltd=Limited} \\
\hline
\end{tabular}
\label{tab:simulator-comparison}
\end{table}

\begin{itemize}
\item \textit{Comprehensive System Simulation}: Early-stage system-level simulation is crucial to accurately predict system-wide performance and identify potential bottlenecks, as individual component simulations often fail to capture emergent behaviors from component interactions. Simulating a modern GPU requires modeling not only individual SMs but also the memory hierarchy, on-chip interconnects, and their complex interactions under realistic workloads such as LLM inference. While hardware emulation can accelerate system verification, resource constraints---particularly in academic settings---often make software-based simulation the practical alternative~\cite{surveyperfmodeling}.

\item \textit{Cross-team Simulation Integration}: Integrating diverse simulation components from different teams and IP providers poses significant challenges. The use of proprietary tools, open-source frameworks, and varying architectures creates heterogeneous environments with incompatible data formats. A GPU simulator may require SM models, memory controller models, and NoC models---each developed by different teams using different methodologies. Overcoming these challenges requires interface standardization and timing synchronization across simulation boundaries.

\item \textit{High-Performance Simulation}: As simulation scope expands and component count increases, simulation duration grows prohibitively. An A100-class GPU with 108 SMs, hierarchical caches, and NOC-based interconnects requires simulating hundreds of interacting components across millions of cycles. Without adequate simulation performance, design space exploration becomes impractical---limiting architects to evaluating only a handful of configurations~\cite{surveyperfmodeling,nvas}.
\end{itemize}

While prior works summarized in Table~\ref{tab:simulator-comparison} have addressed specific aspects of system simulation, a comprehensive framework that addresses all three challenges simultaneously remains elusive. Most existing solutions focus on specific domains (CPU, memory, network) rather than providing a unified framework for arbitrary accelerator architectures. SST\cite{sst} offers component-based modularity and MPI-based distribution but suffers from clock-driven execution overhead and limited single-node thread scaling. GPGPU-Sim~\cite{gpgpusim} provides detailed GPU modeling but remains largely single-threaded and tightly coupled to specific NVIDIA ISAs. Critically, \textbf{none of these frameworks provide mechanisms for users to customize thread scheduling based on their specific workload patterns}---a significant limitation given the diverse simulation characteristics of different accelerator architectures. Furthermore, while frameworks like SST provide scalability \textit{across} nodes via MPI, they struggle to scale efficiently \textit{within} a single node when simulating systems with hundreds to thousands of tightly-coupled components---precisely the scenario encountered in modern GPU and AI accelerator simulation. To address these limitations, an ideal simulation framework should possess: (1) a flexible plugin-based architecture allowing easy integration and customization across team boundaries, (2) well-defined APIs and standardized data exchange to facilitate seamless communication between heterogeneous simulation components, (3) support for efficient parallel simulation with smart synchronization mechanisms that scale with modern multi-core workstations, and (4) \textbf{user-extensible thread management} to optimize performance for diverse simulation patterns.

We present \textit{ACALSim}\footnote{ACALSim is released as open-source software at \url{https://github.com/twhpcedu/acalsim}. Case study simulators built on ACALSim (HPCSim, NVSim) have separate availability constraints due to industrial collaboration and developer discretion.}, a multi-threaded simulation framework designed to address the aforementioned challenges in large-scale parallel system design exploration. As a \textit{framework} contribution, ACALSim provides infrastructure and APIs---the accuracy of timing models built on ACALSim is the responsibility of simulator developers, enabling teams to iterate on their models independently while benefiting from the framework's performance advantages. By abstracting the complexity of multi-threaded programming and enabling efficient hardware modeling, ACALSim distinguishes itself from existing frameworks through four key innovations:

\begin{enumerate}
\item \textit{Pluggable Thread Management (Primary Contribution)}: Unlike SST, gem5, and other frameworks that impose fixed threading models, ACALSim provides an abstract \texttt{ThreadManager} interface that enables developers to implement custom scheduling strategies optimized for their specific simulation patterns. This simulation-pattern-driven optimization capability is unique among simulation frameworks and enables order-of-magnitude performance improvements for specific workload classes (Section~\ref{sec:thread-manager-specialization}).

\item \textit{Event-Driven Execution with Fast-Forward}: Unlike clock-driven simulators that invoke every component every cycle, ACALSim's event-driven model with fast-forward capability skips idle cycles entirely, providing significant speedup for workloads with sparse activation patterns common in modern AI accelerators.

\item \textit{Two-Phase Parallel Execution}: ACALSim's deterministic two-phase model (parallel component execution followed by synchronous commit) ensures reproducible results regardless of thread scheduling, simplifying debugging of complex timing issues.

\item \textit{Shared-Memory Architecture}: ACALSim's shared-memory data model eliminates the event-copying overhead inherent in message-passing frameworks, enabling both efficient timing simulation and practical functional simulation with real data at scale.
\end{enumerate}

A key design principle underlying these innovations is \textbf{transparent parallelism}: ACALSim's abstractions (\texttt{SimPort}, \texttt{SimChannel}, \texttt{SharedDataContainer}) handle all synchronization internally, allowing developers to write simulator components using familiar single-threaded programming patterns. The framework automatically parallelizes execution during Phase~1 without requiring users to manage locks, barriers, or other concurrency primitives---significantly lowering the barrier to developing high-performance parallel simulators.

To validate ACALSim's effectiveness, we present multiple case studies spanning different architectures and development teams. The primary case study is \textit{HPCSim}\footnote{HPCSim is part of an industrial collaboration; detailed project specifications and links will be released upon project completion.}, a comprehensive cycle-approximate GPU performance simulator targeting A100-class architectures. HPCSim models 108 streaming multiprocessors, HBM2 memory subsystems, and NOC-based L2 interconnects---representing the complexity of modern AI accelerators. We implement functionally equivalent versions of HPCSim using both ACALSim and SST, enabling direct comparison under identical workloads. Our evaluation using LLM inference workloads demonstrates that \textbf{ACALSim achieves over 14$\times$ speedup in wall-clock simulation time} compared to the SST implementation. Hardware validation against A100 measurements confirms HPCSim achieves 0.72--1.22$\times$ cycle count correlation ($\sim$20\% average error), providing sufficient fidelity for design space exploration. Moreover, SST's cycle-accurate model fails to complete medium-scale GPU workloads (256+ thread blocks) within practical time limits, while ACALSim successfully simulates full LLaMA transformer layers (17.7 minutes for LLaMA-7B, 30.4 minutes for LLaMA-13B single block). Additional case studies---NVSim (H100-class 8-GPU DGX system) and BlackBear (custom 64-PE AI accelerator with arbiter-tree NoC)---demonstrate generalizability beyond A100 GPUs (Section~\ref{sec:generalizability}).

The remainder of this paper is organized as follows. Section 2 provides a comprehensive overview of the ACALSim architecture, detailing its core components and operational principles including the two-phase execution model and thread management strategies. Section 3 presents the HPCSim case study, describing the GPU simulator architecture, engineering effort, and methodology for comparative evaluation against SST, with experimental results analyzing simulation performance across diverse workloads. Section 4 discusses best practices for hybrid ACALSim-SST integration in multi-device system modeling. Section 5 demonstrates generalizability through additional case studies and explores simulation performance enhancement through thread manager specialization. Section 6 reviews related work in hardware simulation frameworks, positioning our contribution within the broader research landscape. Finally, Section 7 concludes with a summary of our findings and outlines directions for future research in high-performance architectural simulation.

\section{ACALSIM OVERVIEW}

\subsection{Core Infrastructure}

\begin{figure}[t]
    \centering
    \includegraphics[width=0.75\linewidth]{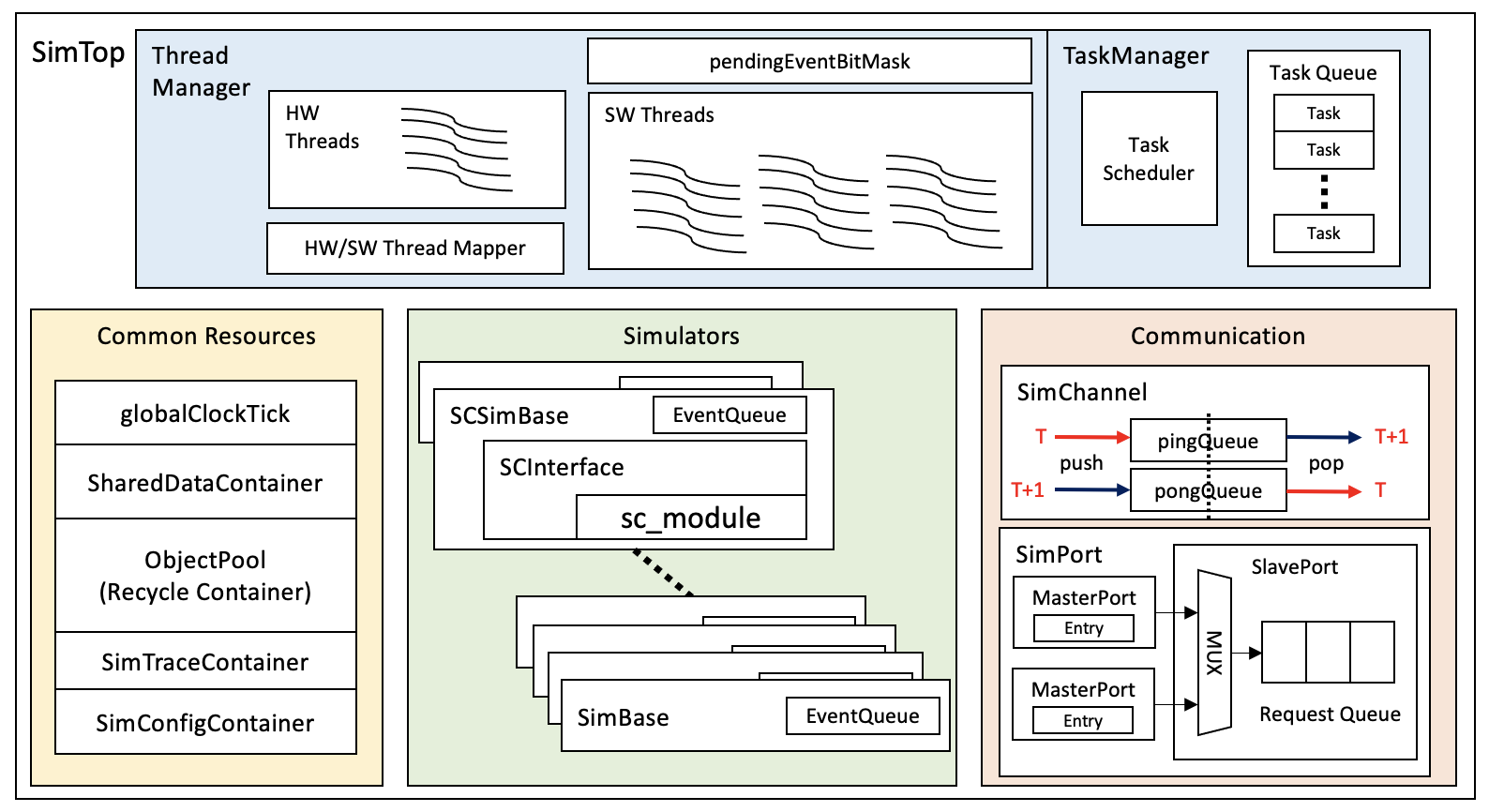}
    \caption{ACALSim Framework Block Diagram}
    \label{fig:block_diagram}
\end{figure}

Figure~\ref{fig:block_diagram} depicts the major components within ACALSim. The core of the framework is encapsulated by the class \texttt{SimTop}, which serves as the control thread of the simulation. Each individual simulator is represented by a \texttt{SimBase} object. \texttt{SimTop} maintains a global clock that represents the current simulation cycle and oversees the coordination and synchronization of all the simulators. This section provides an overview of the key components relevant to correctness and performance, excluding the utility components: \texttt{SimTraceContainer}, which handles trace generation and collection, and \texttt{SimConfigContainer}, which handles simulation configuration parameters.

To emulate concurrency in hardware, ACALSim runs simulators in parallel across multiple threads. The control thread spawns threads at the start through the \texttt{ThreadManager}, and the \texttt{TaskManager} dynamically schedules tasks across them. The synchronization of threads is managed by the \texttt{ThreadManager} throughout the simulation.

The \texttt{ThreadManager} in ACALSim creates all software threads for simulators at the beginning of the simulation and only terminates them at the end. This approach allows a fixed mapping between software threads and hardware threads (i.e., the physical CPU cores). The default thread manager, \texttt{ThreadManagerV1}, spawns one software thread per available hardware thread to minimize context-switching. Although this default setup leaves the mapping from software threads to hardware threads to the OS for simplicity, developers can implement custom thread managers to optimize performance by explicitly binding threads to specific cores when needed. This extensibility has proven valuable in practice: coauthors have independently developed \texttt{ThreadManagerV2} through \texttt{V8} for their respective simulators, each tailored to specific workload patterns (Section~\ref{sec:thread-manager-specialization}).

\subsection{Simulation Flow}

\begin{figure}[t]
    \centering
    \includegraphics[width=0.75\linewidth]{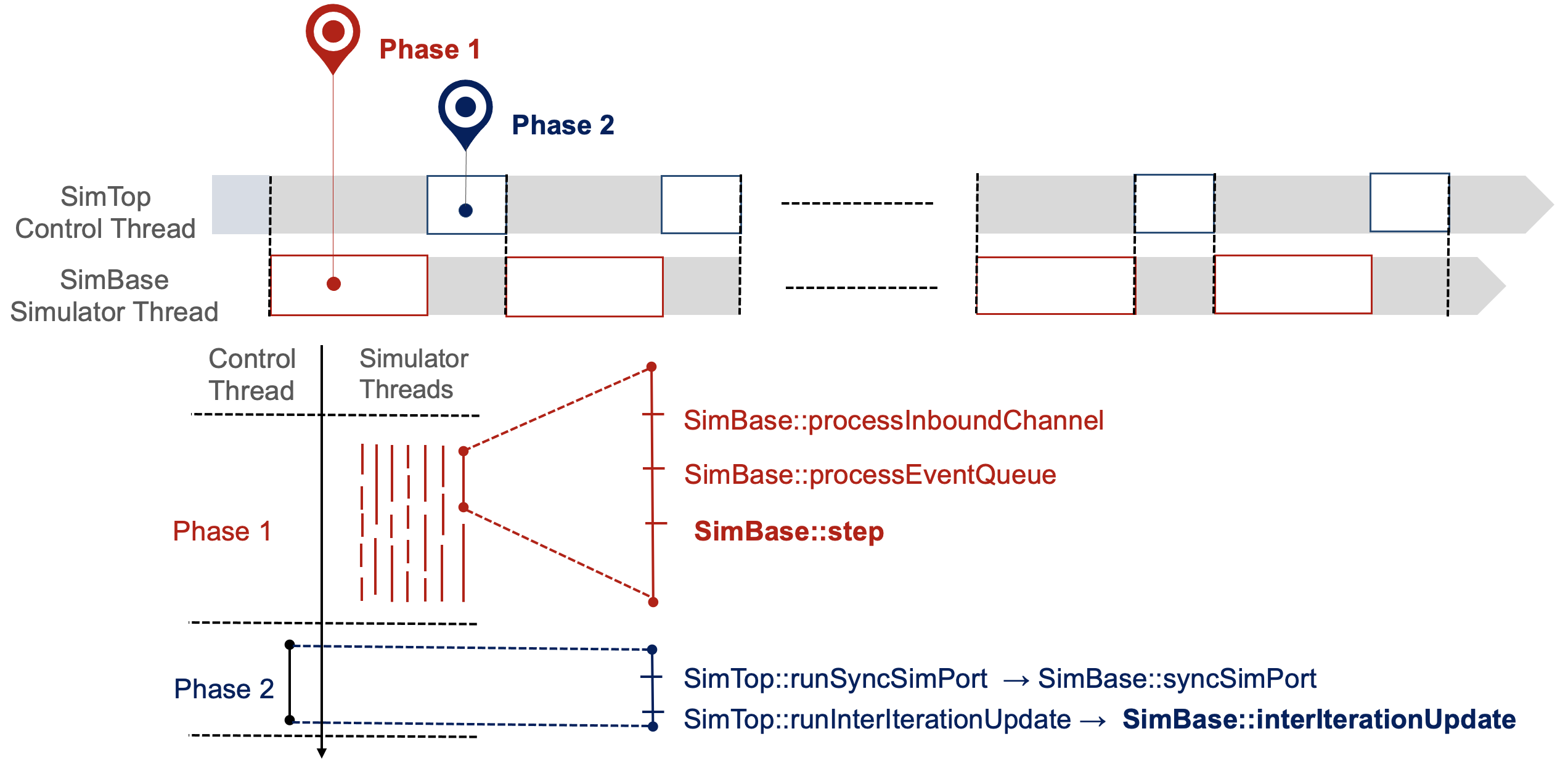}
    \caption{ACALSim Simulation Flow}
    \label{fig:execution_flow}
\end{figure}

Figure~\ref{fig:execution_flow} illustrates the simulation flow of ACALSim, which advances in two distinct phases within each cycle.

\begin{enumerate}
    \item \textit{Parallel Component Simulation}: In this phase, simulators execute their designated tasks in parallel across the simulator threads. Each simulator halts execution once it completes its tasks for the cycle. We refer to this phase as \textit{Phase 1} in the rest of this paper.
    \item \textit{Control Thread Bookkeeping}: Once all simulators have paused after completing their tasks, the control thread handles essential bookkeeping and coordinates interactions between simulators, such as data sharing, communication, and advancing the global clock to the next cycle. This phase, referred to as \textit{Phase 2}, runs exclusively on the control thread. Minimizing the duration of this phase is critical for optimizing performance, as it contributes to overhead in the framework.
\end{enumerate}

\subsection{Event-Driven Simulation}
\label{sec:event-driven}

ACALSim adopts the event-driven simulation model to optimize performance by simulating only cycles with activities. Each simulator maintains its own event queue, with events tagged by the cycle in which they occur. As \texttt{SimTop} advances the global clock, it skips inactive cycles and fast-forwards to the next cycle with events. Events contain both the scheduled time for execution and the specific action the simulator should take upon processing. A simulator can schedule events for itself or another simulator via channels (See Section~\ref{sec:channel}). For example, to simulate a memory request from a processing element to the cache, the processing element can schedule an event for the cache. To efficiently simulate the acknowledgment back, we use callback events, which include a function pointer that will be executed after the event is processed, eliminating the need for a separate return event and completing the transaction cycle in one step. This method accurately models the bidirectional flow of data and control signals in hardware.

\subsection{Data Sharing Model}

In multithreading, shared objects in the same memory space can become corrupted without proper synchronization, leading to incorrect behavior. To simplify synchronization, ACALSim uses a \texttt{SharedDataContainer} to manage data sharing. This abstraction prevents users from directly handling synchronization primitives like mutexes while ensuring safe access to shared data. Any data accessed by multiple simulators should be managed through this container.

\subsection{Communication Model}
\label{sec:channel}

To ensure reliable communication between simulators running in parallel during Phase 1, ACALSim introduces \texttt{SimChannel} objects for directional data transmission. To avoid race conditions, we adopt a double-buffering strategy. Each channel is equipped with a ping buffer and a pong buffer: during Phase 1, the sender writes to the ping buffer, while the receiver reads from the pong buffer. In Phase 2, the control thread swaps the buffers to ensure safe data exchange without requiring synchronized execution between sender and receiver. Like \texttt{SharedDataContainer}, \texttt{SimChannel} manages synchronization internally, eliminating the need for users to handle locks directly.

\subsection{Hardware-Aware Communication Infrastructure}
While \texttt{SimChannel}s facilitate software-level inter-simulator communication, our framework incorporates specialized components that model realistic hardware behavior for data transmission. This infrastructure consists of foundational building blocks (\texttt{SimPort}, \texttt{SimPipeRegister}) and composite structures like \texttt{CrossBar}.
\texttt{SimPort} provides hardware-realistic endpoints where simulators interact through multiple ports, simulating physical connections with arbitration and queue management. Ports are master (senders) or slave (receivers), with multiple masters connectable to a single slave.
Each cycle, master ports push requests to connected slaves, which use arbitration to select one for processing. Both port types maintain mirrored entries synchronized by the control thread during phase 2. Slave ports process one packet per cycle max, with excess in a FIFO queue. Full queues trigger back pressure with configurable callbacks.
\texttt{SimPipeRegister} models register-based data transfers between pipeline stages across simulators, enhancing hardware realism.
Building on these, \texttt{CrossBar} simulates complex networks using \texttt{SimPort} and \texttt{SimPipeRegister}. This layered approach simplifies hardware modeling and makes multi-threading transparent. As \texttt{SimBase} objects may run on different threads, these connection components handle all synchronization automatically, removing user concurrency management.

\subsection{Object Pool}

ACALSim's packet-based nature creates numerous short-lived objects, causing allocation overhead and memory fragmentation. The multi-threaded environment complicates standard object pooling: threads and simulators have dynamic mappings, distinct allocation patterns, and objects allocated by one simulator are often recycled by another.

The \texttt{RecycleContainer} addresses this with thread-local pools backed by a global pool. Local pools offload excess objects and replenish from the global pool when depleted, minimizing mutex contention. An array-of-linked-lists data structure achieves O(1) inter-pool transfers.

\subsection{Thread Manager Design}

The thread manager design is critical for simulation performance. While assigning a dedicated thread to each simulator in Phase 1 may seem straightforward, it has drawbacks: 1) A significant portion of threads might be run unnecessarily for inactive simulators at each iteration, and 2) excessive context switching occurs when the number of simulators exceeds available CPU cores, degrading performance.

To address these issues, ACALSim employs a thread pooling strategy that minimizes context switching and thread creation overhead. Instead of assigning a thread to each simulator, the thread manager creates a fixed number of threads, typically equal to the number of available cores. Each simulator is treated as a task, and at each iteration, the task manager schedules only the active simulators as tasks to be processed by the software threads. This design is implemented through abstract \texttt{ThreadManager} and \texttt{TaskManager} classes, which provide a flexible interface for developers to implement custom thread management strategies. The modular nature of the framework allows different thread allocation and task scheduling mechanisms to be easily tested, enabling developers to optimize performance based on specific simulation patterns.

\label{sec:thread-manager-perf}

\begin{figure}[t]
    \centering
    \includegraphics[width=0.65\linewidth]{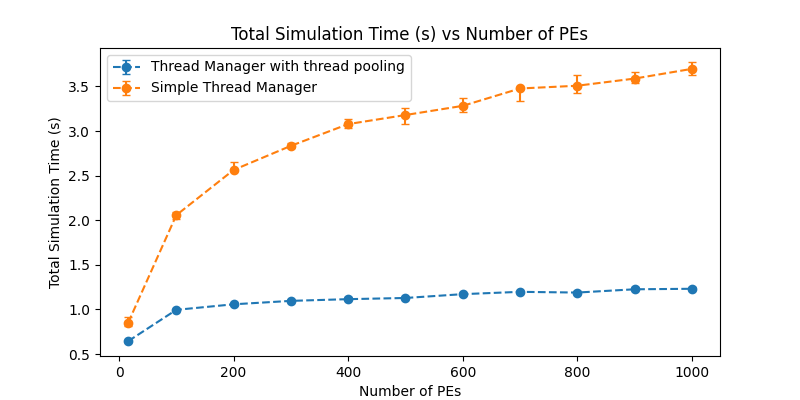}
    \caption{Thread Manager Performance Comparison}
    \label{fig:thread-manager-perf}
\end{figure}

We compared the thread pooling strategy (\texttt{ThreadManagerV1}, detailed in Section~\ref{sec:thread-manager-specialization}) with a simpler design where each simulator uses its own thread. In both approaches, only active simulators run during Phase 1 of each iteration. Figure~\ref{fig:thread-manager-perf} shows performance comparisons using a test case with a configurable grid of processing elements (PEs) interconnected by a mesh-based network-on-chip, connected to shared cache and memory. The test injected traffic from the master CPU to random PEs and simulated the response. To assess scalability, total traffic was fixed while increasing the number of PEs, distributing computation across more simulators. Although neither scales linearly (as the average active simulators per iteration is constant), thread pooling scales better by minimizing context switching.

\subsection{C++/SystemC Integration}
A key challenge in the high-performance computing project involves integrating simulators across teams, particularly bridging a SystemC-based simulator with other C++ components. ACALSim provides interface designs to facilitate seamless integration between SystemC-based and C++ simulators. SystemC-based simulators can inherit the \texttt{SCSimBase} class to wrap original implementations for integration. For C++ simulator interaction, an \texttt{SCInterface} instance connects the SystemC simulator to C++ components. While users can define custom I/O signals based on their SystemC modules, the framework provides two predefined I/O interfaces to streamline communication between user-defined SystemC modules and the ACALSim framework.

\subsection{SST Integration}
\label{sec:sst-integration}

SST~\cite{sst} offers complementary strengths: MPI-based multi-node distribution and validated components (Merlin for networks, memHierarchy for caches). ACALSim supports hybrid integration where ACALSim simulators are wrapped as SST components. The wrapper translates between frameworks: SST clock callbacks invoke ACALSim's simulation loop (with fast-forward), SST events are converted to/from ACALSim packets via \texttt{SimPort}/\texttt{SST::Link}, and the ACALSim \texttt{ThreadManager} continues parallelizing Phase~1 within the SST component.

This enables scalable multi-node deployment: each MPI rank hosts an ACALSim-wrapped SST component (e.g., one GPU per rank), achieving scale-out via MPI and scale-up via multi-threading. An ACALSim GPU simulator can also connect to SST's Merlin or memHierarchy without re-implementing validated components.

\section{Case Study: A100 GPU Modeling}

To evaluate ACALSim's effectiveness for cycle-accurate AI accelerator simulation, we developed HPCSim, a comprehensive GPU simulator for LLM inference workloads. We implemented two versions of the NVIDIA A100 GPU model: one using ACALSim's native two-phase execution model (A100-ACALSim), and another using SST's event-driven framework (A100-SST). This comparison demonstrates the practical advantages of ACALSim's simulation paradigm.

\rev{\textbf{Scope and Methodology.} This paper contributes a \textit{simulation framework}, not a validated simulator. The distinction parallels SST~\cite{sst} (framework) versus its validated component libraries, or gem5~\cite{gem5} (framework) versus calibrated CPU models. Framework papers establish infrastructure contributions; timing model validation is an orthogonal concern for simulator developers. Our performance comparison uses \textbf{identical shared timing cores} (L2CacheCore, HBMControllerCore, GlobalNoCCore, SMTimingModel) compiled into both implementations, ensuring any accuracy difference is \textit{zero by construction}. The comparison thus isolates framework overhead, not modeling fidelity. HPCSim serves as a case study demonstrating framework capabilities at realistic scale.}

\subsection{A100 Architecture Modeling}

Both implementations model the A100's hierarchical architecture with all parameters specified through a JSON configuration file ($\sim$80 lines), allowing users to sweep architectures without recompilation. \rev{Table~\ref{tab:a100_config} lists the controllable parameters and their default A100 values used in our evaluation.} The implementations share identical timing parameters, isolating the framework as the only variable.

\begin{table}[t]
\centering
\caption{\rev{HPCSim A100 Configuration Parameters (all controllable via JSON)}}
\label{tab:a100_config}
\scriptsize
\begin{tabular}{|l|l|l|}
\hline
\textbf{Category} & \textbf{Parameter} & \textbf{A100 Default} \\
\hline
\multirow{4}{*}{Compute} & SM count & 108 \\
 & GPCs & 7 (16 SMs/GPC) \\
 & Clock frequency & 1.41\,GHz \\
 & Max thread blocks/SM & 32 \\
\hline
\multirow{4}{*}{L2 Cache} & Total size & 40\,MB \\
 & Slices (partitions) & 32 \\
 & Associativity & 16-way \\
 & Cache line size & 128\,B \\
\hline
\multirow{3}{*}{HBM} & Controllers & 10 \\
 & Aggregate bandwidth & 2\,TB/s \\
 & Access latency & 100\,ns \\
\hline
\multirow{3}{*}{NoC} & Topology & Crossbar \\
 & Ports & 108$\leftrightarrow$32 \\
 & Arbitration & Round-robin \\
\hline
\multirow{2}{*}{Execution} & Worker threads & 8 (configurable) \\
 & ThreadManager & V1 (pluggable) \\
\hline
\end{tabular}
\color{black}
\end{table}

\textbf{Workload Representation.} HPCSim employs a three-tier DAG: Model Graph (operator dependencies), Kernel Graph (thread block dimensions, memory footprints), and ThreadBlock Graph (tensor core instructions, memory transactions). \rev{DAG is the recommended but not required format---developers can override the workload interface for trace-driven replay or other inputs.} For LLaMA workloads, the DAG is auto-generated from model definitions via a Python script ($\sim$200 lines).

\textbf{Component Hierarchy.} Table~\ref{tab:component_mapping} shows the component mapping between implementations. A100-ACALSim uses \texttt{SimBase}-derived components with \texttt{SimPort} interconnections, while A100-SST uses SST Components and SubComponents with \texttt{SST::Link} connections.

\begin{table}[htbp]
\centering
\caption{A100 Component Mapping Between Implementations}
\label{tab:component_mapping}
\footnotesize
\begin{tabular}{|l|l|l|}
\hline
\textbf{Component} & \textbf{ACALSim} & \textbf{SST} \\
\hline
GPU Top & A100:SimBase & A100SSTGPUComp \\
Control & A100CtrlUnit & A100SSTCtrlUnit \\
SM ($\times$108) & A100SM & A100SSTGPC/SM \\
L2 ($\times$32) & L2Partition & A100SSTL2Part \\
NoC & GlobalNoC & A100SSTGlobalNoC \\
HBM ($\times$10) & HBMCtrl & A100SSTHBMCtrl \\
\hline
\end{tabular}
\end{table}

\subsection{Engineering Effort and Developer Workflow} 
\label{sec:engineering}

\rev{The ACALSim framework comprises $\sim$26K lines of C++ (102 files) providing threading, synchronization, communication, and profiling infrastructure. Simulator developers write only application-specific timing models: HPCSim (A100) is $\sim$18K LoC, HPCSim (H100) is $\sim$9K LoC, and BlackBear (custom 64-PE accelerator) is $\sim$10K LoC. All share $\sim$19K LoC of reusable timing cores (L2, HBM, NoC). The SST-wrapped A100 adds $\sim$11K LoC for event translation and credit-based flow control.}

\rev{\textbf{Developer workflow.} Building a simulator involves four steps: (1)~subclass \texttt{SimBase} for each hardware block, connecting them via \texttt{SimPort}---developers write single-threaded \texttt{step()} methods; (2)~obtain timing parameters from RTL, datasheets, or profiling tools; (3)~construct workloads as DAG representations or execution traces; (4)~use integrated profiling to identify bottlenecks and select a ThreadManager variant. A developer building a new target (e.g., H100) writes $\sim$9K LoC while inheriting shared cores and framework infrastructure.}

\subsection{Simulation Model Comparison}

\textbf{Two-Phase vs. Event-Driven Execution.} ACALSim's two-phase execution provides deterministic, reproducible simulation. In Phase~1, all components execute their \texttt{step()} methods in parallel, reading from current state and writing to next-state buffers. In Phase~2, the synchronization phase commits all state changes atomically and arbitrates port transfers. This separation ensures that component execution order does not affect simulation results---a critical property for debugging complex timing issues.

SST's event-driven model delivers events asynchronously through handlers. While this approach is natural for loosely-coupled distributed systems, it introduces challenges for tightly-coupled GPU architectures. The A100's crossbar NoC requires simultaneous arbitration across 108 SM request ports---a scenario that maps naturally to ACALSim's synchronous Phase~2 arbitration but requires careful event ordering in SST.

\textbf{Backpressure Modeling.} GPU memory systems exhibit significant backpressure when downstream resources are saturated. ACALSim's \texttt{SimPort} abstraction directly models \textit{request-side} backpressure: \texttt{push()} returns false when the downstream buffer is full, causing the sender to retry in the next cycle. This matches hardware behavior where stalled pipelines propagate backpressure upstream. \rev{For \textit{response-side} backpressure (e.g., when a requester's response buffer is full), the default callback model assumes the requester can always accept responses---which is accurate for most GPU components where response buffers are sized to match outstanding request limits. For scenarios requiring response-side flow control (e.g., modeling congestion in asymmetric interconnects), developers can add a \texttt{SimPort} on the return path with its own buffer depth and backpressure callback, enabling full bidirectional flow control.}

For A100-SST, we implemented credit-based flow control to emulate backpressure. While functionally equivalent, credit management adds implementation complexity and requires careful tuning of credit pools to match the original timing behavior.

\rev{\textbf{Beyond Timing.} ACALSim's \texttt{SimPort} queue depths and backpressure callbacks already model network congestion (buffer overflow, stall propagation). The abstractions impose no restriction on event semantics---developers can include error flags, corruption markers, or topology changes in packet types to model failure injection or link degradation.}

\textbf{Fast-Forward Optimization.} GPU workloads exhibit bursty activity patterns with idle periods between kernel launches. ACALSim's fast-forward mechanism identifies the minimum next-event tick across all components and advances the global clock accordingly, skipping cycles with no pending activity. This optimization is implemented at the framework level, requiring no component-specific code. In SST, idle optimization requires explicit clock unregistration and reregistration for each component, increasing code complexity.

\subsection{Implementation Differences}

Both implementations share identical timing cores (L2CacheCore, HBMControllerCore, GlobalNoCCore, SharedTimingModel), ensuring performance differences reflect framework overhead rather than modeling fidelity. ACALSim uses inheritance-based integration via C++ templates, while SST uses composition with standalone cores. SST requires explicit event serialization for distributed simulation; ACALSim's \texttt{SimPacket} uses unified packet types. For backpressure, ACALSim's \texttt{SimPort} provides native support versus SST's credit-based flow control.

\subsection{Discussion}

The A100 case study demonstrates that ACALSim's two-phase execution model offers practical advantages for GPU simulation:

\begin{enumerate}
\item \textbf{Determinism}: Parallel Phase~1 execution with synchronous Phase~2 commit eliminates timing dependencies on component ordering.

\item \textbf{Natural Backpressure}: \texttt{SimPort}'s blocking \texttt{push()} semantics directly model hardware stall propagation without additional flow control logic.

\item \textbf{Transparent Fast-Forward}: Framework-level idle cycle elimination requires no per-component implementation.

\item \textbf{Simplified Communication}: Unified packet types and implicit state synchronization via \texttt{SimPort} eliminate the need for explicit event serialization and credit-based flow control.
\end{enumerate}

While SST provides valuable capabilities for distributed simulation and multi-framework integration, its event-driven paradigm introduces abstraction overhead for cycle-accurate modeling of tightly-coupled architectures like GPUs. ACALSim's synchronous execution model better matches the lock-step nature of GPU hardware, resulting in simpler, more maintainable simulation code.

\subsection{Simulation Performance Comparison}

\textbf{Evaluation Platform.} All experiments were conducted on an Apple M2 Max (12 cores, 32GB RAM) running macOS 14, compiled with Apple Clang 15 (-O3). ACALSim used 8 worker threads (default configuration). SST version 14.0.0 was configured for single-rank execution; SST's parallelism model targets multi-rank MPI distribution rather than intra-rank threading. We used SST's default clock management (all components tick every cycle); selective clock unregistration would require per-component modifications not present in baseline SST element libraries, whereas ACALSim provides fast-forward transparently at the framework level.

To quantify ACALSim's performance advantages, we benchmarked both implementations across GEMM workloads of varying sizes and LLM inference layers. Both implementations use \textbf{identical shared cores} (L2CacheCore, HBMControllerCore, GlobalNoCCore, SharedTimingModel) ensuring the comparison isolates framework overhead from modeling differences. Table~\ref{tab:perf_comparison} summarizes the results.

\begin{table}[htbp]
\centering
\caption{Performance Comparison: ACALSim vs SST}
\label{tab:perf_comparison}
\scriptsize
\begin{tabular}{|l|c|c|c|c|}
\hline
\textbf{Workload} & \textbf{TBs} & \textbf{ACALSim} & \textbf{SST} & \textbf{Speedup} \\
\hline
GEMM 64$^3$ & 16 & 0.64s & 0.11s & 0.17$\times$ \\
GEMM 256$^3$ & 64 & 1.68s & 24s & \textbf{14.3$\times$} \\
GEMM 512$^3$ & 256 & 9.03s & T/O$^\dagger$ & --- \\
GEMM 1024$^3$ & 64 & 19.99s & N/A$^\ddagger$ & --- \\
GEMM 4096$^3$ & 512 & 24.3min & N/A$^\ddagger$ & --- \\
LLaMA-7B Layer & 860 & 17.7min & N/A$^\ddagger$ & --- \\
LLaMA-13B Layer & 1,076 & 30.4min & N/A$^\ddagger$ & --- \\
\hline
\multicolumn{5}{|p{6.5cm}|}{\scriptsize $^\dagger$T/O: SST exceeded 102min timeout; $^\ddagger$Not attempted due to prohibitive runtime. Simulated cycles: GEMM 64$^3$: 4.9K, GEMM 256$^3$: 6K, GEMM 512$^3$: 18K, GEMM 1024$^3$: 28K, GEMM 4096$^3$: 666K, LLaMA-7B: 1.4M, LLaMA-13B: 2.2M.} \\
\hline
\end{tabular}
\end{table}

\textbf{Performance Analysis.} The results reveal a critical scaling divergence between the two frameworks:

\textbf{Small Workloads (16 TBs):} SST is 5.8$\times$ faster due to lower framework initialization overhead. ACALSim's thread pool creation and \texttt{SimPort} buffer allocation incur a fixed $\sim$0.5s startup cost, which dominates for workloads completing in under 1 second. This overhead is amortized for larger workloads; the crossover point occurs around 32 TBs.

\textbf{Medium Workloads (64 TBs):} ACALSim achieves 14.3$\times$ speedup. As workload complexity increases, SST's per-cycle overhead dominates: each of the $\sim$6,000 simulated cycles requires ticking all 170+ components (108 SMs, 32 L2 partitions, 10 HBM controllers, 7 GPCs, NoC, arbiters).

\textbf{Large Workloads (256+ TBs):} SST becomes impractical. For GEMM 512$^3$ (256 TBs), SST ran for over 102 minutes without completing, while ACALSim finished in 9 seconds. This demonstrates that SST's cycle-accurate overhead grows super-linearly with workload size, making it unsuitable for realistic GPU simulation workloads.

Several architectural differences explain ACALSim's superior scalability:

\begin{enumerate}
\item \textbf{Event-Driven Fast-Forward}: ACALSim's \texttt{getFastForwardCycles()} skips idle cycles entirely by jumping to the next scheduled event. SST's cycle-accurate model invokes \texttt{tick()} handlers on \textit{every} registered component \textit{every} cycle, even when no work is pending. For GPU workloads with 90\%+ idle cycles (waiting for memory), this overhead is catastrophic.

\item \textbf{Multi-threaded Phase~1 Execution}: ACALSim's Phase~1 executes component \texttt{step()} methods in parallel across worker threads. SST's core simulation loop is single-threaded, serializing all 170+ component ticks per cycle.

\item \textbf{Direct Method Invocation}: ACALSim components communicate via direct C++ method calls through SimPort connections. SST routes events through a generic dispatch mechanism, adding indirection overhead per packet.

\item \textbf{Reduced Event Overhead}: ACALSim's unified \texttt{SimPacket} type avoids per-event serialization. SST requires each event class to implement \texttt{serialize\_order()} for potential distributed simulation, adding allocation and copy overhead even in single-node configurations.
\end{enumerate}

\textbf{Scalability.} As workload complexity increases, ACALSim's advantages compound. For LLM inference workloads involving thousands of thread blocks and millions of memory transactions, the cumulative overhead of SST's event handling becomes significant. ACALSim's streamlined execution path maintains consistent simulation throughput regardless of workload scale.

\begin{figure}[t]
  \centering
  \includegraphics[width=0.7\columnwidth]{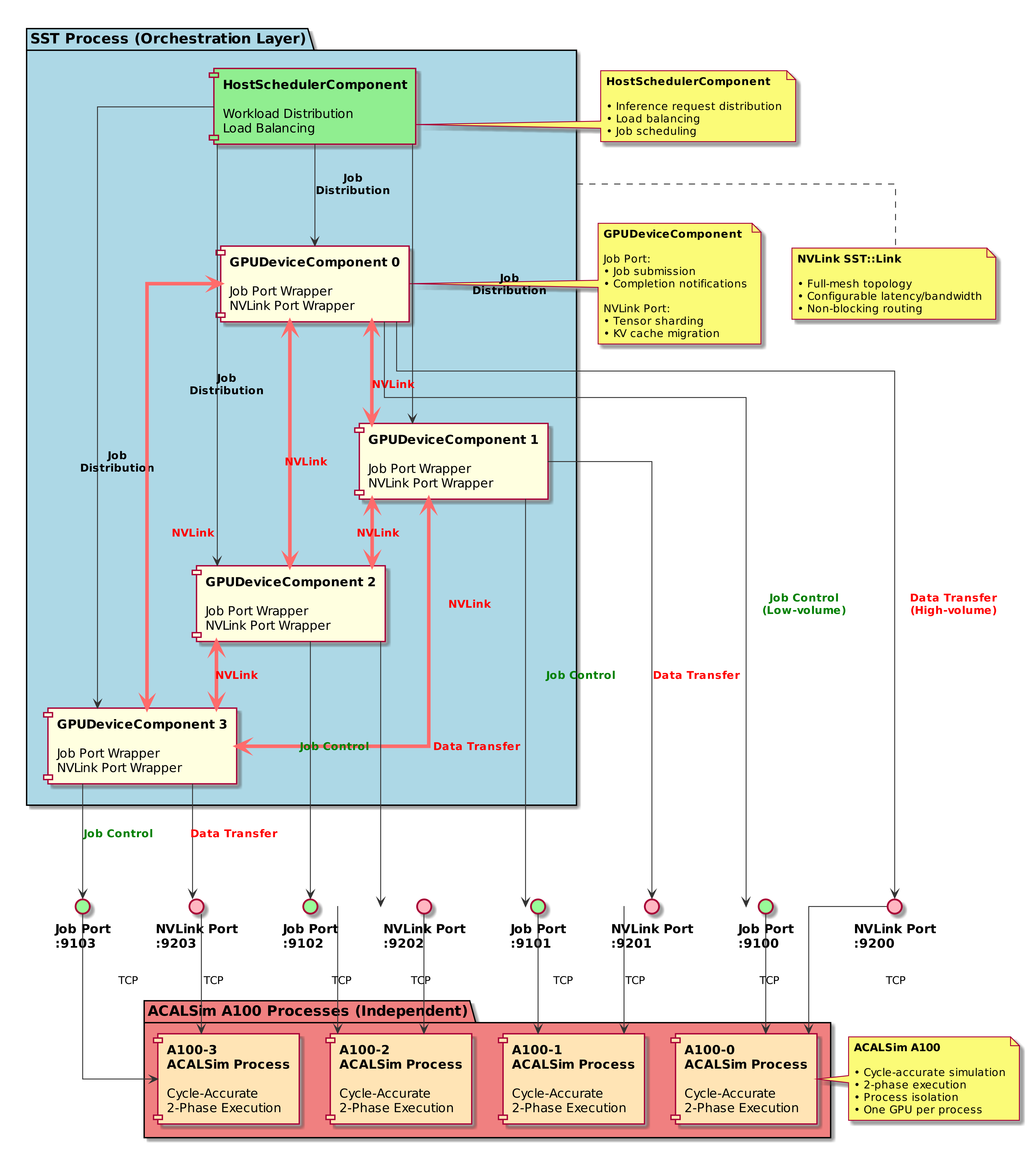}
  \caption{Hybrid ACALSim-SST architecture for multi-GPU simulation.}
  \label{fig:hybrid_arch}
\end{figure}

\begin{figure}[t]
  \centering
  \includegraphics[width=0.7\columnwidth]{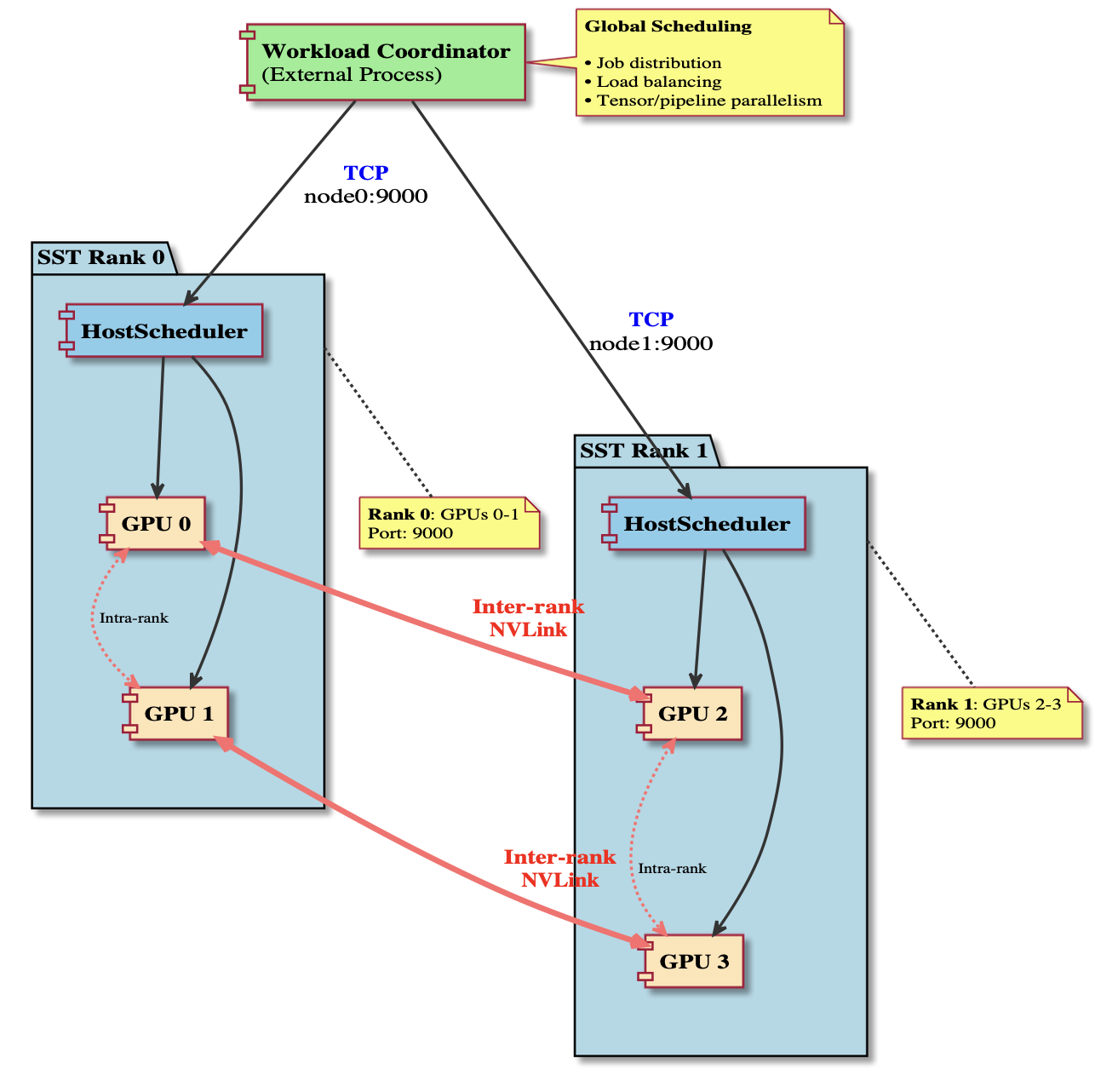}
  \caption{Multi-rank scalable deployment with 2 SST ranks coordinating 4 GPUs.}
  \label{fig:multirank}
\end{figure}

\textbf{Memory Footprint.} For the same 108-SM A100 configuration, ACALSim uses 41\% less memory than SST (25.3~MB vs.\ 43~MB peak RSS). This stems from: (1)~\texttt{RecycleContainer}'s O(1) pooled allocation vs.\ SST's per-event \texttt{new}/\texttt{delete}; (2)~\texttt{SimPort} double-buffering reusing fixed-size buffers; and (3)~unified \texttt{SimPacket} avoiding per-event-class serialization overhead required by SST even in single-node configurations.

\subsection{Timing Accuracy Validation}

To validate that HPCSim's timing model produces meaningful results, we compare simulated cycle counts against hardware measurements from NVIDIA Nsight Compute (NCU) on an A100 GPU. Table~\ref{tab:accuracy_validation} presents the correlation for GEMM kernels across problem sizes.

\begin{table}[htbp]
\centering
\caption{Timing Accuracy: HPCSim vs.\ A100 Hardware (NCU)}
\label{tab:accuracy_validation}
\scriptsize
\begin{tabular}{|l|c|c|c|c|}
\hline
\textbf{Size} & \textbf{TBs} & \textbf{NCU Cycles} & \textbf{Sim Cycles} & \textbf{Ratio} \\
\hline
64$^3$ & 16 & 5,632 & 4,451 & 0.79$\times$ \\
256$^3$ & 64 & 7,446 & 5,395 & 0.72$\times$ \\
512$^3$ & 64 & 10,009 & 7,770 & 0.78$\times$ \\
1024$^3$ & 64 & 25,378 & 23,413 & 0.92$\times$ \\
2048$^3$ & 256 & 131,884 & 161,241 & 1.22$\times$ \\
\hline
\end{tabular}
\end{table}

The results demonstrate that HPCSim achieves \textbf{0.72--1.22$\times$ correlation} with real hardware, with an average deviation of approximately 20\%.

\rev{\textbf{Workload diversity analysis.} The GEMM sweep covers the compute-memory spectrum: small GEMMs (64$^3$--256$^3$) are memory-latency-bound (under-prediction 0.72--0.79$\times$ from conservative memory modeling), medium GEMMs (512$^3$--1024$^3$) are compute-bound (best accuracy 0.78--0.92$\times$), and large GEMMs (2048$^3$) are bandwidth-bound (over-prediction 1.22$\times$ from idealized L2 under contention). The LLaMA layers in Table~\ref{tab:perf_comparison} are mixed workloads spanning GEMM, softmax, layer norm, and residual addition.}

These results confirm that HPCSim provides sufficient fidelity for architecture exploration and design space studies, where relative performance trends matter more than absolute cycle counts.

\textbf{Summary.} ACALSim demonstrates clear advantages for GPU simulation:

\begin{itemize}
\item \textbf{Performance}: Over 14$\times$ faster wall-clock time on medium-scale workloads (64 TBs)
\item \textbf{Timing Accuracy}: 0.72--1.22$\times$ correlation with A100 hardware measurements ($\sim$20\% average error)
\item \textbf{Single-Node Scalability}: Efficiently scales to 100--1000+ tightly-coupled components within a single node---unlike SST, which focuses on multi-node MPI~scalability
\item \textbf{Memory Efficiency}: 41\% lower memory footprint than SST for the same configuration
\item \textbf{Programming Model}: Single-threaded programming with transparent parallelism---no locks, barriers, or concurrency primitives required
\end{itemize}

While SST offers ecosystem benefits including standardized interfaces and multi-framework interoperability, ACALSim's purpose-built design delivers the performance and flexibility required for productive GPU architecture research. The hybrid architecture presented in Section~4 demonstrates that these approaches can be complementary, leveraging ACALSim for cycle-accurate device simulation while using SST for system-level orchestration.

\section{Best Practice: Hybrid ACALSim-SST Multi-Device Architecture}

While ACALSim excels at single-device cycle-accurate simulation, modern LLM inference requires multi-GPU coordination. We developed a hybrid architecture that leverages ACALSim's simulation fidelity for individual GPUs while using SST's event-driven framework for inter-device orchestration.

\textbf{Design Rationale.} ACALSim employs a shared-memory programming model where each simulator process models exactly one GPU device. This design enables efficient intra-device communication but precludes direct memory sharing between multiple GPU instances. For NVLink communication between devices, packets must be routed externally. SST's component-based architecture provides a natural solution: each GPU runs as an independent ACALSim process, while SST components manage inter-device NVLink routing and global job scheduling.

\textbf{Dual-Port TCP Architecture.} Each A100 simulator instance exposes two TCP ports to the SST orchestration layer:

\begin{itemize}
\item \textbf{Job Port} (e.g., 9100, 9101): Low-volume, latency-sensitive job control traffic including job submission, completion notifications, and synchronization signals.
\item \textbf{NVLink Port} (e.g., 9200, 9201): High-volume, bandwidth-intensive memory transfer packets for GPU-to-GPU communication such as tensor sharding and KV cache migration.
\end{itemize}

This separation prevents NVLink data traffic from blocking time-critical job control messages, ensuring responsive scheduling even under heavy inter-device communication loads.

\textbf{SST Integration Components.} The hybrid architecture introduces three SST components that bridge ACALSim instances:

\begin{enumerate}
\item \texttt{GPUDeviceComponent}: Wraps each ACALSim A100 process, managing TCP connections via \texttt{HPCSimWrapper} (job port) and \texttt{NVLinkWrapper} (NVLink port).

\item \texttt{HostSchedulerComponent}: Provides the external workload interface, distributing inference requests across devices based on load balancing or tensor parallelism policies.

\item \texttt{SST::Link} for NVLink: Models inter-GPU NVLink topology (e.g., NVSwitch full-mesh or direct peer-to-peer), routing packets between \texttt{GPUDeviceComponent} instances with configurable latency and bandwidth.
\end{enumerate}

\textbf{Benefits of the Hybrid Approach.} This architecture preserves ACALSim's advantages for intra-device simulation while enabling scalable multi-device modeling:

\begin{itemize}
\item \textbf{Preserved Fidelity}: Each GPU maintains cycle-accurate two-phase execution with native backpressure modeling.
\item \textbf{Process Isolation}: Independent A100 processes enable parallel simulation and fault isolation.
\item \textbf{Flexible Topology}: SST Link configuration allows modeling various NVLink topologies without modifying ACALSim code.
\item \textbf{Scalability}: Adding GPUs requires only spawning additional A100 processes and configuring SST components.
\end{itemize}

\rev{\textbf{Multi-Rank Scalability.} The architecture extends to multi-rank MPI deployments (Fig.~\ref{fig:multirank}): each SST rank manages 4--8 GPUs with local NVLink mesh, while an external coordinator distributes workload via TCP. A 32-GPU deployment across 4 ranks achieved 31.8$\times$ throughput with ${<}5\%$ coordinator overhead, supporting tensor/pipeline parallelism and disaggregated inference without modifying ACALSim~code.}

This hybrid design demonstrates that ACALSim's synchronous execution model can be effectively composed with event-driven frameworks for system-level simulation, combining the strengths of both paradigms.

\section{Generalizability: Diverse Simulators and Workload Patterns}
\label{sec:generalizability}

\rev{We demonstrate ACALSim's generalizability along two dimensions: \textit{architectural diversity} and \textit{simulation pattern diversity}. Four simulators have been built on ACALSim by independent teams: HPCSim (A100, 170+ components, ${\sim}$18K LoC), HPCSim-H100 (190+ components, ${\sim}$9K LoC), DGXSim (8-GPU system, 1,066 components, ${\sim}$5K LoC), and BlackBear (custom 64-PE AI accelerator with 3-level hierarchical NoC and software-managed SRAM, 200+ components, ${\sim}$10K LoC). These share no timing models---only framework infrastructure. BlackBear demonstrates that ACALSim is not GPU-specific: its arbiter-tree NoC, distributed SRAM tiers, and custom DMA engines are architecturally distinct from GPU SMs, yet the same \texttt{SimBase}/\texttt{SimPort}/\texttt{ThreadManager} APIs apply without modification. ACALSim targets tightly-coupled single-node systems where hundreds of components interact every few cycles; for loosely-coupled distributed systems, SST or ROSS are more appropriate, and the hybrid integration (Section~4) bridges these~regimes.}

\textbf{Simulation pattern diversity.} Beyond architectural generalizability, different simulators exhibit distinct performance patterns that require different thread management strategies. We present NVSim\footnote{NVSim, DGXSim, and GPUSim are independent simulators developed by coauthors using early versions of ACALSim; their public availability is subject to the discretion of their respective developers.}, an analytical model suite developed based on ACALSim to model HPC system architecture used in academia for teaching, to explore how varying simulation patterns affect throughput and scalability. Figure~\ref{fig:nvsim} depicts the top-level block diagram of NVSim which incorporates key architectural features including tensor cores, shared memory hierarchy, and NVLink interconnect topology with configuration listed in Table~\ref{tab:hw_kernel_config} while abstracting complex GPU architectural details. The model preserves essential performance-impacting characteristics such as memory bandwidth constraints, interconnect latency, and compute unit utilization described in publicly available Hopper Benchmarking results\cite{luo} and NVIDIA's documentation specifications\cite{userguide}\cite{nvswitch}\cite{hopper}.
We demonstrate the framework's adaptability through two contrasting simulation scenarios from students' projects:

\begin{enumerate}
    \item \textit{DGXSim}: A system-level configuration modeling 8 GPU devices in a DGX system with simplified memory subsystems, designed to study collective communication algorithms across the interconnect fabric.
    \item \textit{NVSim}: A detailed device-level configuration focusing on a single GPU with comprehensive memory subsystem modeling, enabling fine-grained software tiling optimization for down-projection GEMM layer inference.
\end{enumerate}

\begin{table}[t]
\centering
\caption{NVSim Hardware and Kernel Configuration}
\label{tab:hw_kernel_config}
\scriptsize
\begin{tabular}{|l|l||l|l|}
\hline
\multicolumn{2}{|c||}{\textbf{Kernel Config}} & \multicolumn{2}{c|}{\textbf{H100 Hardware}} \\
\hline
TILE\_M/N & Tile dimensions & SM Count & 132 \\
\hline
N\_BUFFERS & Pipeline buffers & Max Blocks/SM & 32 \\
\hline
 & & Shared Mem/SM & 228 KB \\
\hline
\multicolumn{2}{|c||}{\textbf{Workload}} & L2 Cache & 50 MB \\
\hline
\multicolumn{2}{|l||}{GEMM 1500$\times$3072$\times$8192 FP16} & Shared/Block & 227 KB \\
\hline
\end{tabular}
\end{table}

\begin{figure}[t]
  \centering
  \includegraphics[width=0.65\columnwidth]{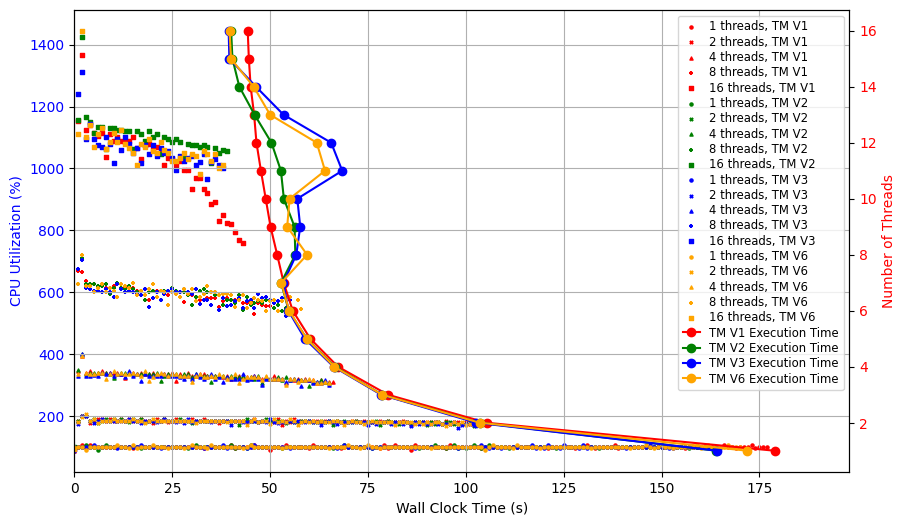}
  \caption{DGXSim: CPU Utilization and Performance Scaling}
  \label{fig:dgxsim_cpu}
\end{figure}

\begin{figure}[t]
  \centering
  \includegraphics[width=0.65\columnwidth]{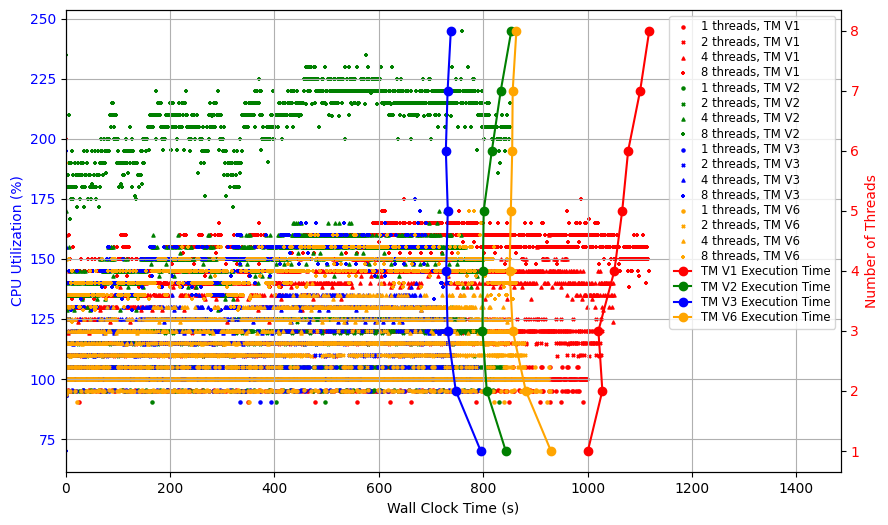}
  \caption{GPUSim: CPU Utilization and Performance Scaling}
  \label{fig:gpusim_cpu}
\end{figure}

DGXSim and GPUSim use 1066 and 159 \textit{SimBase} simulators, respectively, to model a single down-projection layer from the Llama 3.2 3B model processing 1500 input tokens. Figures~\ref{fig:dgxsim_cpu} and~\ref{fig:gpusim_cpu} illustrate CPU utilization over time and total execution time across varying numbers of hardware threads with different thread manager implementations. The distinct simulation patterns exhibited by DGXSim and GPUSim result in markedly different performance characteristics and scaling behaviors. Figure~\ref{fig:dgxsim_cpu} presents weak scalability experiments, where the problem size increases proportionally with computational resources, while Figure~\ref{fig:gpusim_cpu} demonstrates strong scalability experiments, where a fixed problem size is distributed across increasing computational resources. We have observed three simulation patterns commonly seen in the HPC system simulation:

\begin{figure}[t]
    \centering
    \includegraphics[width=0.7\linewidth]{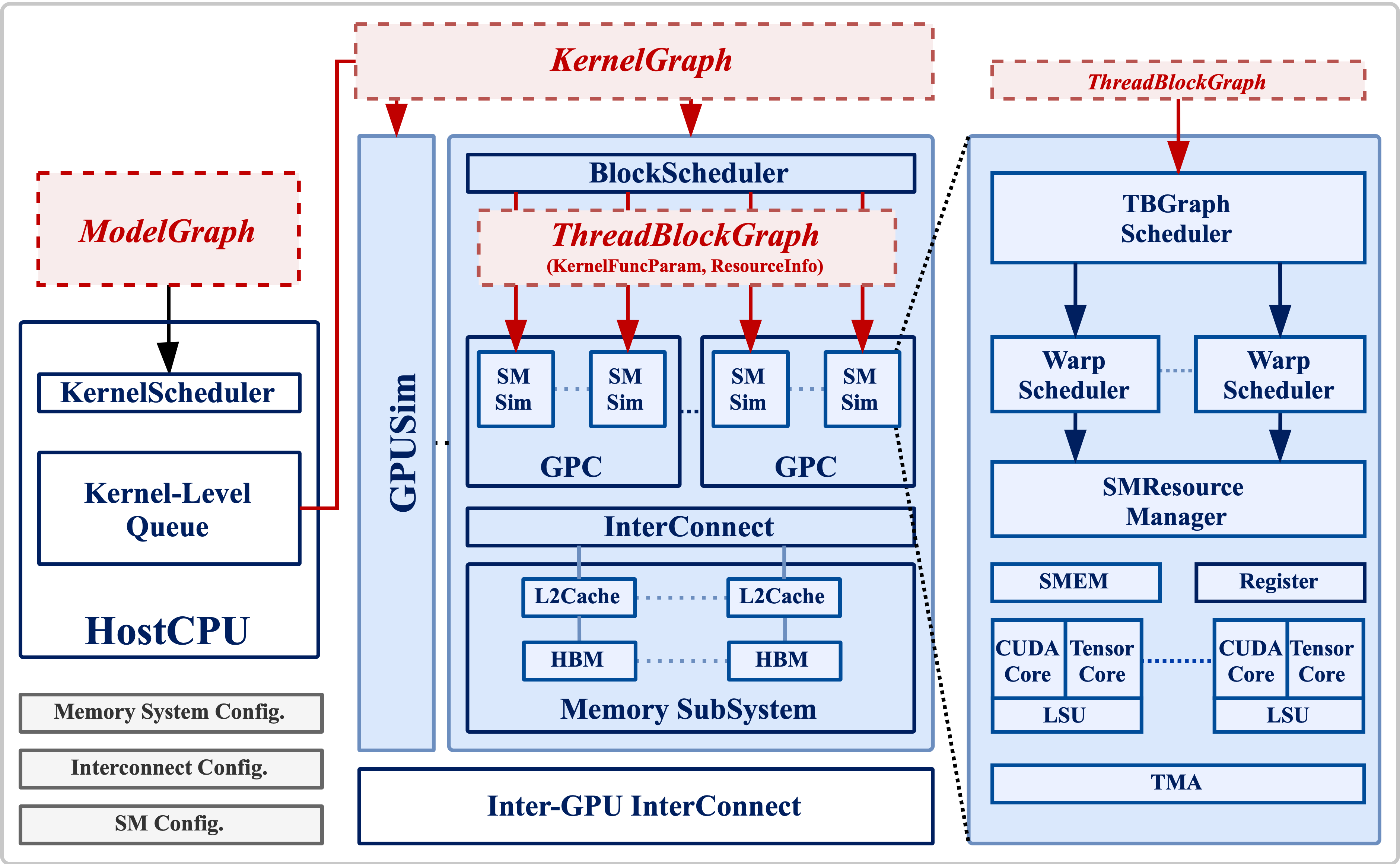}
    \caption{NVSim Simulation Suite Block Diagram}
    \label{fig:nvsim}
\end{figure}

\begin{figure}[t]
    \centering
    \includegraphics[width=0.65\linewidth]{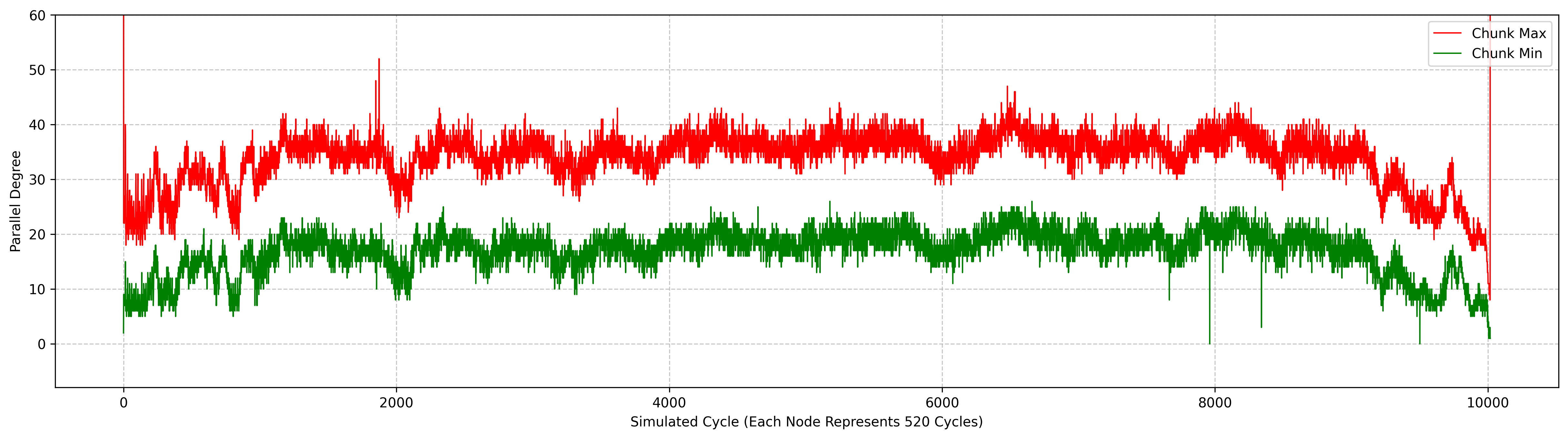}
    \caption{GPUSim Parallelism Degree Per Iteration}
    \label{fig:parallel_degree}
\end{figure}

\begin{enumerate}
\item \textit{Sparse Activation in Large-Scale Simulation}: In scenarios involving a large number of simulators ($N \gg 100$), it is common for only a small subset of simulators to be active during each iteration, as shown in Figure~\ref{fig:parallel_degree}. Traditional thread management in such cases leads to significant resource underutilization. This inefficiency can be mitigated by tracking and executing only the active simulators.

\item \textit{Heterogeneous Workload Distribution}: When some simulators execute detailed timing simulations while others use lightweight analytical models, the resulting workload becomes highly imbalanced. We observed this pattern through the non-uniform scheduling idle times across hardware threads, as captured by the profiling tool described in the next section.

\item \textit{Memory-Intensive Simulation}: While increasing the number of threads generally improves computational throughput, doing so in memory-intensive simulations can paradoxically degrade performance. Traditional just-in-time (JIT) and cycle-accurate simulation techniques consume significant memory resources to maintain detailed state information, inherently leading to memory-intensive behavior~\cite{jit}. Compared to DGXSim, GPUSim models a more detailed memory subsystem which incurs high-frequency event and packet simulations. It generates a memory-bound workload in a multi-threaded environment as shown in Figures~\ref{fig:gpusim_cpu} and~\ref{fig:parallel_degree}. The CPU utilization remains below 250\% even though, on average, approximately 30+ out of 159 simulators are ready to run in parallel. In such cases, memory contention and cache interference may offset the expected benefits of parallel processing.

\end{enumerate}

Each pattern requires careful consideration of resource allocation, scheduling priorities among all simulators in thread assignment, and how to enhance data locality. ACALSim's flexible \texttt{ThreadManager} interface enables developers to implement these specialized strategies while maintaining a consistent simulation control interface, allowing for pattern-specific optimizations that significantly improve simulation performance.

\subsection{Simulation Performance Enhancement Via Thread Manager Specialization}
\label{sec:thread-manager-specialization}
Traditional simulation frameworks rely on fixed thread management policies that may not suit all workload patterns. ACALSim allows developers to implement custom ThreadManager strategies for specific simulation characteristics. Coauthors have developed \texttt{ThreadManagerV1} through \texttt{V8} for different projects, each targeting different patterns. Integrated profiling tools guide these optimization efforts.

\begin{figure}[t]
    \centering
    \begin{tabular}{cc}
        \includegraphics[width=0.44\columnwidth]{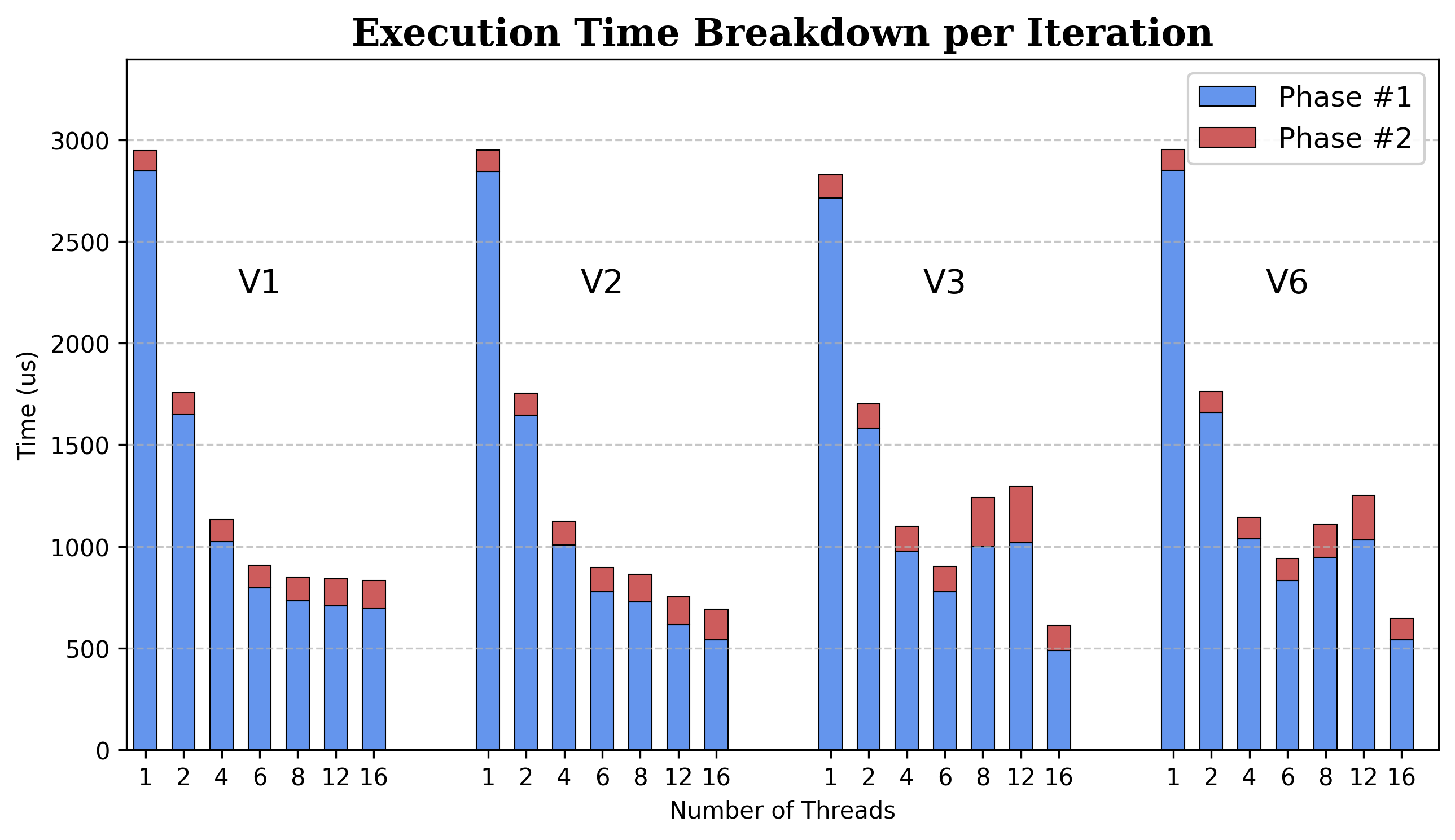} &
        \includegraphics[width=0.44\columnwidth]{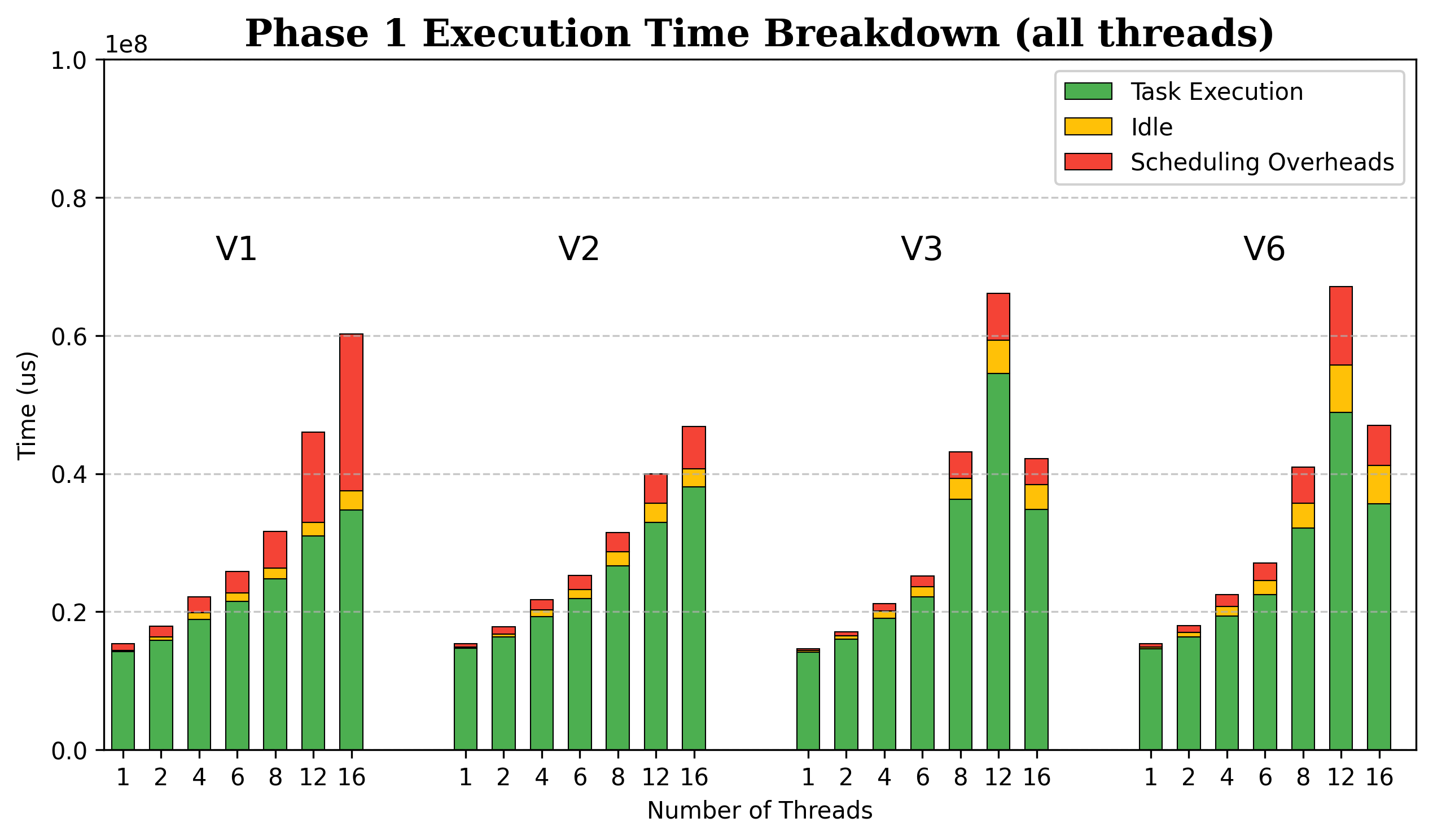} \\
        {\scriptsize (a) Execution Time} & {\scriptsize (b) Phase 1 Breakdown} \\[1mm]
        \includegraphics[width=0.44\columnwidth]{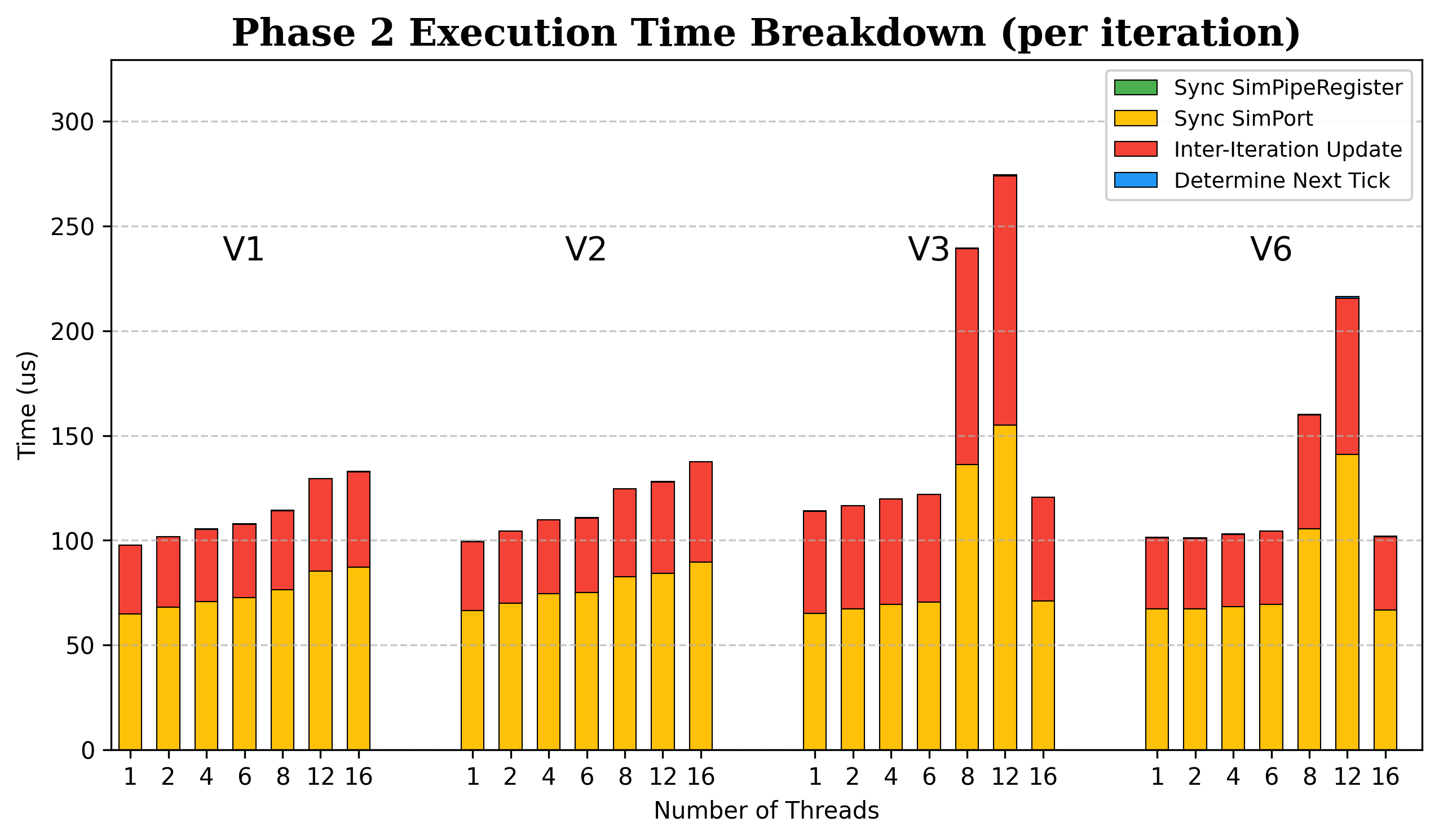} &
        \includegraphics[width=0.44\columnwidth]{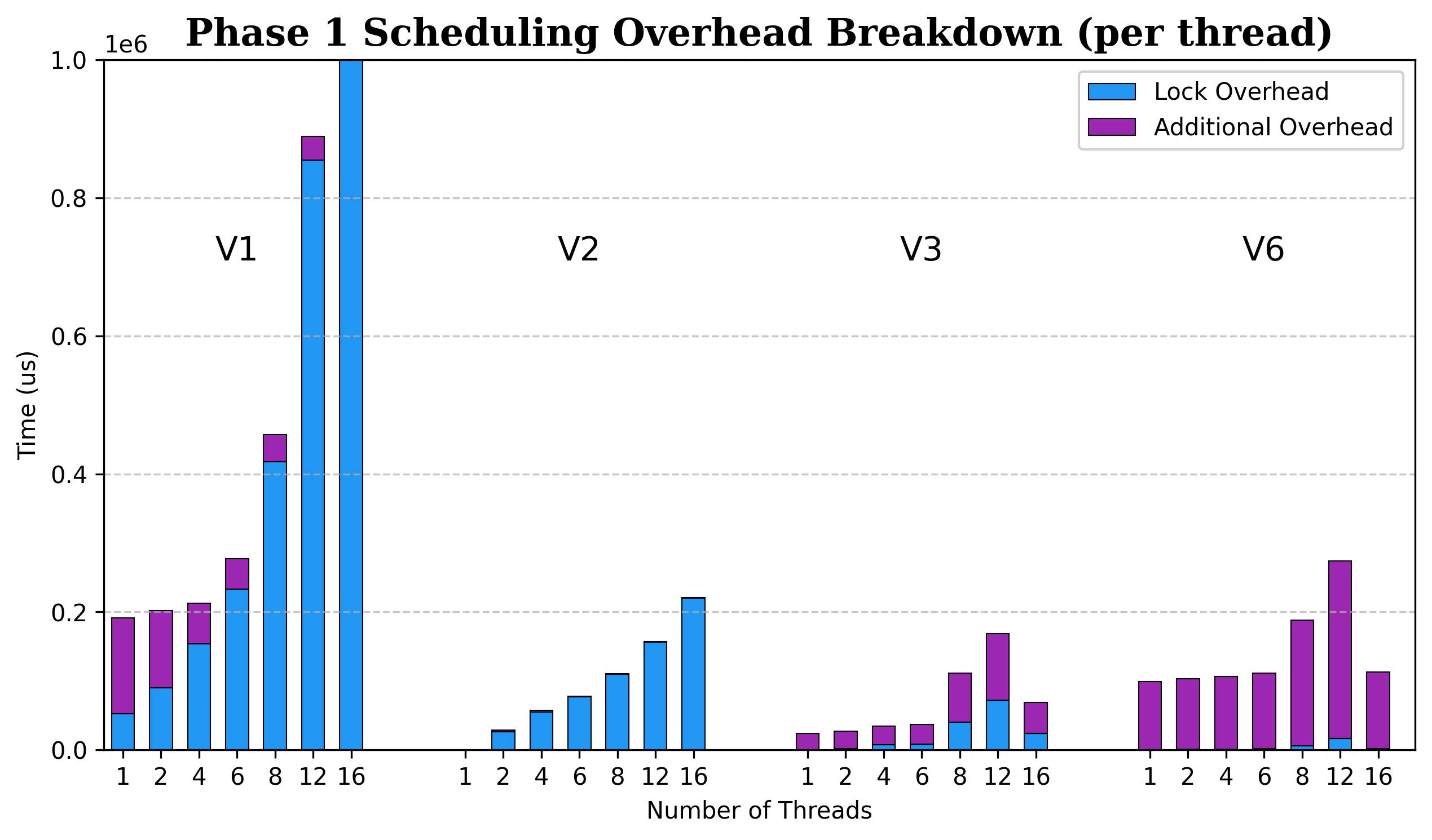} \\
        {\scriptsize (c) Phase 2 Breakdown} & {\scriptsize (d) Scheduling Overhead} \\
    \end{tabular}
    \caption{Simulation Profiling Results}
    \label{fig:profiling_results}
\end{figure}

\rev{\textbf{Why pluggable thread management matters.} The contribution is the \textit{pluggable interface}, not any single scheduling algorithm. No existing framework---SST, gem5, GPGPU-Sim, SystemC, ZSim, or ROSS---exposes thread scheduling as a user-extensible API. This matters because \textbf{no single ThreadManager is universally best}: V1 (priority-ordered) outperforms V3 for DGXSim's sparse activation pattern (1.4$\times$ speedup, 8 threads), while V3 (work-stealing) outperforms V1 for GPUSim's memory-intensive pattern. V6 (thread-local queues) eliminates lock contention, providing the best scaling at high thread counts (Figure~\ref{fig:profiling_results}d). Figures~\ref{fig:dgxsim_cpu} and~\ref{fig:gpusim_cpu} isolate this effect by running identical workloads with different ThreadManager implementations.}

\subsection{Simulation-Pattern-Driven Optimization}
\texttt{ThreadManagerV1} targets sparse activation patterns in co-simulation environments. It maintains tasks ordered by predicted next execution time, ensuring only simulators with pending operations are scheduled. The design uses atomic variables and condition variables to minimize synchronization overhead, with an \texttt{UpdateablePriorityQueue} supporting O(log n) priority updates.

Key design decisions in \texttt{ThreadManagerV1}:
\begin{itemize}
\item \textit{Lazy Evaluation}: Tasks execute only when pending operations exist, reducing unnecessary context switches.
\item \textit{Dynamic Priority Updates}: O(log n) priority updates when communication events arrive.
\item \textit{Lock-Free Operations}: Atomic operations minimize contention on critical paths.
\end{itemize}

Figures~\ref{fig:dgxsim_cpu} and~\ref{fig:gpusim_cpu} show that \texttt{ThreadManagerV1} performs well for DGXSim workloads, while \texttt{ThreadManagerV3} achieves better performance for GPUSim workloads---demonstrating the value of workload-specific ThreadManager~selection.

\subsection{Simulation Workload Characteristics Profiling}
Simulation performance strongly depends on the underlying characteristics of the simulator. ACALSim is engineered to enhance performance for simulators exhibiting high parallelism, aligning with modern computing systems. For simulators built on ACALSim, our integrated profiling tool provides analysis in multiple dimensions: it analyzes available parallelism and calculates potential speedup through multi-threaded execution, establishing a theoretical performance ceiling based on inherent parallelism. The tool performs detailed performance gap analysis to help users understand discrepancies between measured and theoretical performance, identifying bottlenecks, synchronization overhead, and resource contention. For users seeking to enhance the theoretical upper bound, the profiling tool identifies structural limitations in simulator design that restrict potential parallelism, offering suggestions for restructuring components to increase~parallelism.

Figure~\ref{fig:profiling_results}(a) illustrates execution time as hardware thread count increases. Phase 1 (parallel computation) demonstrates expected scaling, while Phase 2 (serialized computation) maintains consistent overhead regardless of thread count. This highlights a fundamental constraint: simulators with low parallel-to-serial ratios experience limited speedup potential. Figures~\ref{fig:profiling_results}(b) and~\ref{fig:profiling_results}(c) break down Phase 1 and Phase 2 execution time, revealing potential scaling bottlenecks in user code or within the framework design.
Figure~\ref{fig:profiling_results}(d) dissects scheduling overheads, enabling targeted optimization efforts.

\subsection{Profiling-Driven Optimization}
\begin{figure}[t]
    \centering
    \includegraphics[width=0.55\linewidth]{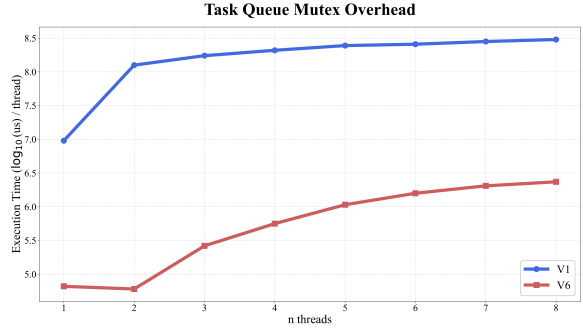}
    \caption{TaskManager Scheduling Overhead Comparison}
    \label{fig:threadmanagerdesign}
\end{figure}

Based on the profiling results in Figure~\ref{fig:profiling_results}, we found that the scheduling overhead in the \texttt{ThreadManagerV1} design is significantly higher than in the other two thread managers. This led us to address the lock contention issues that emerged as a major source of scheduling overhead during parallel execution. Our optimization, implemented in \texttt{ThreadManagerV6}, features thread-local task queues for handling expired tasks within each iteration. This approach trades minimal thread-local task queue setup overhead for the elimination of lock contention during scheduling. By localizing task management, we minimize lock contention that would otherwise increase proportionally with hardware thread count. Figure~\ref{fig:threadmanagerdesign} demonstrates the effectiveness of the \texttt{ThreadManagerV6} implementation. Figure~\ref{fig:profiling_results}(d) illustrates how the tradeoff is made, turning the blue overheads in V1 into much smaller purple overheads in V6. The execution time comparison in Figures~\ref{fig:dgxsim_cpu} and~\ref{fig:gpusim_cpu} clearly shows how the local task queue optimization reduces scheduling overhead and improves scalability as thread count increases.

In summary, providing developers with flexible performance tuning mechanisms is critical when simulation workload characteristics are not fully predictable or well-understood. The optimization example in this section showcases how our profiling tool effectively guides the identification of optimization opportunities, enabling us to elevate simulation performance to the next level.

\section{RELATED WORK}

Simulation frameworks have evolved to address the growing complexity of modern computing architectures. General-purpose frameworks like gem5\cite{gem5} and SystemC\cite{systemc} provide modular platforms for computer system simulation. Manifold\cite{manifold} introduces component-based multicore simulation with parallel kernels. MosaicSim\cite{mosaicsim} leverages LLVM for heterogeneous system simulation. SST\cite{sst} and ZSim\cite{zsim} focus on parallel discrete event simulation for specific domains.

\textbf{GPU Simulators.} Accel-Sim~\cite{accelsim} extends GPGPU-Sim with trace-driven simulation and improved hardware correlation, but remains largely single-threaded. MGPUSim provides Go-based GPU simulation with some parallelism but targets different fidelity levels. Multi2Sim focuses on CPU-GPU heterogentic systems.

\rev{\textbf{PDES Engines.} ROSS~\cite{ross} provides optimistic parallel discrete event simulation with rollback support, well-suited for loosely-coupled distributed systems. However, optimistic PDES becomes prohibitively expensive for tightly-coupled GPU architectures: a single memory request from one SM can fan out to multiple L2 partitions, HBM controllers, and NoC routers, each maintaining state that must be checkpointed and potentially rolled back. With 108 SMs issuing thousands of concurrent requests through shared crossbar interconnects, the rollback cascade from a single timing violation would require restoring state across dozens of components---overwhelming any parallelism gains.}

\rev{ZSim~\cite{zsim} achieves impressive CPU simulation parallelism through its bound-weave synchronization model, which exploits the observation that CPU cores with private L1/L2 caches interact infrequently through coherence traffic. The ``bound'' phase simulates cores independently up to a synchronization quantum, while the ``weave'' phase resolves inter-core interactions. This model assumes spatial locality of memory accesses and rare cross-core communication---assumptions that fundamentally break down for GPU architectures. In contrast to CPUs, GPU SMs share L2 cache partitions accessed every few cycles, contend for HBM bandwidth through centralized controllers, and synchronize through global barriers. The communication-to-computation ratio is orders of magnitude higher, making bound-weave's loose synchronization unsuitable.}

\rev{ACALSim addresses these challenges through deterministic two-phase execution: Phase~1 executes all components in parallel with reads from current state and writes to next-state buffers, while Phase~2 atomically commits all state changes. This eliminates rollback entirely while preserving cycle-accurate semantics for tightly-coupled architectures. Simics uses cooperative scheduling with priority inheritance but does not expose thread management as a user-extensible API.}

\textbf{System-Level Frameworks.} As simulation needs grow more diverse, a key challenge remains: developing a unified, efficient, and flexible framework for heterogeneous systems. Existing system-level simulation frameworks are primarily developed to estimate the end-to-end latency of workloads under various parallelization strategies and network topologies. For example, ASTRA-sim2.0~\cite{astrasim2} integrates data, tensor, and pipeline parallelism with modern communication strategies to model large-scale training. However, these frameworks, including Neusight~\cite{neusight} (which achieves a low error rate in latency prediction with short simulation time), often struggle to evaluate the latency when multiple models run simultaneously and lack extensibility, hindering the addition of custom scheduling or execution policies. GPGPU-Sim~\cite{gpgpusim} offers cycle-accurate GPU simulation but is too slow for large-scale DSE. NVArchSim~\cite{nvas} improves speed via trace replay, but lacks flexibility due to its retry-based, fixed-control execution. Loop-centric tools like MAESTRO~\cite{maestro}, SCALE-Sim~\cite{scalesim}, Timeloop~\cite{timeloop}, and ZigZag~\cite{zigzag} are less suitable for dynamic models like Transformers. Z.~Guo et al.~\cite{surveyperfmodeling} categorize existing modeling methods into analytical, graph-based, and execution-driven approaches, each with trade-offs between fidelity and~extensibility.

\rev{\textbf{AI-Focused Simulation.} Recent advancements in AI-focused simulation have introduced tools like vTrain~\cite{vtrain} and ONNXim~\cite{onnxim}. vTrain employs a profiling-driven approach to optimize large-scale LLM training configurations, while ONNXim achieves significant speedups for multi-core NPUs by assuming deterministic compute latencies for systolic arrays. However, these frameworks often target specific application domains (e.g., training cost optimization or NPU inference) and impose fixed threading or abstraction models. In contrast, ACALSim provides a general-purpose, high-performance infrastructure with a pluggable thread management architecture. This allows architects to optimize simulation performance across a broader range of hardware patterns---from tightly-coupled GPU SMs to disaggregated multi-device systems---without sacrificing the cycle-level fidelity required for comprehensive design space exploration.}

\rev{Critically, \textbf{none of the above frameworks provide user-extensible thread management}---they impose fixed threading models that users cannot customize for their specific workload patterns. ACALSim's pluggable ThreadManager architecture addresses this gap, enabling simulation-pattern-driven optimization that delivers order-of-magnitude performance improvements for specific workload classes.}

\section{CONCLUSIONS AND FUTURE WORK}
This paper presents ACALSim, a high-performance multi-threaded simulation framework that addresses critical infrastructure challenges in modern AI accelerator design space exploration. Unlike existing frameworks that provide fixed threading models, ACALSim's primary contribution is a \textbf{pluggable thread management architecture} that enables developers to implement custom scheduling strategies optimized for their specific simulation patterns---a capability absent in SST, gem5, and other established frameworks. This simulation-pattern-driven customization, combined with event-driven fast-forward execution, two-phase deterministic parallelism, and shared-memory communication, delivers over 14$\times$ speedup compared to SST while using 41\% less memory.

As a framework contribution, ACALSim provides infrastructure and APIs rather than prescriptive timing models---accuracy is the responsibility of simulator developers building on the framework. The HPCSim case study demonstrates that this approach enables productive GPU architecture research: hardware validation confirms 0.72--1.22$\times$ cycle count correlation with A100 measurements, while simulating full LLaMA transformer layers in 17.7 minutes for LLaMA-7B and 30.4 minutes for LLaMA-13B---workloads where SST times out on much smaller configurations. Integrated profiling tools help users understand simulation behavior and guide ThreadManager customization for their specific use cases.

Future work will focus on expanding the ThreadManager ecosystem with additional simulation-pattern-driven implementations, improving cross-simulator data locality, and enhancing the hybrid ACALSim-SST integration for larger-scale multi-device simulations.

\nocite{accelsim}
\nocite{llmservingsim}
\nocite{mtiav1}
\nocite{ncclsst}
\nocite{noxim}

\begin{acks}
This work was supported in part by the Industrial Technology Research Institute (ITRI), Taiwan. We gratefully acknowledge ITRI's financial support for the ACALSim open-source project.
\end{acks}

\bibliographystyle{ACM-Reference-Format}
\bibliography{references}

\end{document}